\newif\ifsingle

\newif\ifFullVersion
\FullVersiontrue 

\ifsingle
\documentclass[11pt,draftclsnofoot, onecolumn]{IEEEtran}		
\else		
\documentclass[10pt,final, twocolumn]{IEEEtran}
\fi

\usepackage{KalmanNet}
\usepackage{STSP}
\title{KalmanNet: Neural Network Aided Kalman Filtering for Partially Known Dynamics
}
\author{  
	\IEEEauthorblockN{Guy Revach, Nir Shlezinger, Xiaoyong Ni, Adrià López Escoriza, Ruud J. G. van Sloun, and Yonina C. Eldar\\ 
	} 
	\thanks{ 
	  Parts of this work focusing on linear Gaussian state space models were \textcolor{NewColor}{presented} at the IEEE International Conference on Acoustics, Speech, and Signal Processing (ICASSP) 2021 \cite{revach2021kalmannet}.
		G. Revach, X. Ni and A. L. Escoriza are with the Institute for Signal and Information Processing (ISI), D-ITET, ETH Zürich, Switzerland, 
		(e-mail: grevach@ethz.ch; xiaoni@student.ethz.ch; alopez@student.ethz.ch). 
		N. Shlezinger is with the School of ECE, Ben-Gurion University of the Negev, Beer Sheva, Israel (e-mail: nirshl@bgu.ac.il). 
		R. J. G. van Sloun is with the EE Dpt., Eindhoven University of Technology, and with Phillips Research, Eindhoven,  The Netherlands (e-mail: r.j.g.v.sloun@tue.nl). 
		Y. C. Eldar is with the Faculty of Math and CS, Weizmann Institute of Science, Rehovot, Israel (e-mail: \mbox{yonina.eldar@weizmann.ac.il}). 
	} 
}
\vspace{-0.75cm}

\begin{document}
	
	\maketitle
	\pagestyle{plain}
	\thispagestyle{plain}
%
%
\begin{abstract}
State estimation of dynamical systems in \acl{rt} is a fundamental task in signal processing. For systems that are well-represented by a fully known \acl{lg} \ac{ss} model, the celebrated \ac{kf} is a low complexity optimal solution. However, both linearity of the  underlying \ac{ss} model and accurate knowledge of it are often not encountered in practice. Here, we present \acl{kn}, a \acl{rt} state estimator that learns from data to carry out Kalman filtering under non-linear dynamics with partial information. By incorporating the structural \ac{ss} model with a dedicated \acl{rnn} module in the flow of the \ac{kf}, we retain data efficiency and interpretability of the classic algorithm while implicitly learning complex dynamics from data. We demonstrate numerically  that \acl{kn} overcomes non-linearities and model mismatch, outperforming classic filtering methods operating with both mismatched and accurate domain knowledge. 
\end{abstract}
\acresetall 
%
%
\section{Introduction}\label{sec:Intro}
\textcolor{NewColor}{Estimating the hidden state of a dynamical system from noisy observations in \acl{rt} is one of the most fundamental tasks in signal processing and control, with applications in localization, tracking, and navigation \cite{durbin2012time}. In a pioneering work from the early 1960s \cite{kalman1960new, kalman1961new, kalman1963new}, based on work by {Wiener} from 1949 \cite{wiener1949extrapolation}, Rudolf Kalman introduced the \ac{kf},  a \ac{mmse} estimator that is applicable to time-varying systems in \acl{dt}, which are characterized by a linear \ac{ss} model with \ac{awgn}.}
The low-complexity implementation of the \ac{kf}, combined with its sound theoretical basis, resulted in it quickly becoming the leading workhorse of \acl{se} in systems that are well described by \ac{ss} models in \acl{dt}. The \ac{kf} has been applied to problems such as radar target tracking \cite{gruber1967approach}, trajectory estimation of ballistic missiles \cite{larson1967application}, and estimating the position and velocity of a space vehicle in the Apollo program \cite{mclean1962optimal}. 

While the original \ac{kf} assumes linear \ac{ss} models, many problems encountered in practice are governed by \acl{nl} dynamical equations. Therefore, shortly after the introduction of the original \ac{kf}, \acl{nl} variations of it were proposed, such as the \ac{ekf} \cite{gruber1967approach, larson1967application} and the \ac{ukf} \cite{julier1997new}. \textcolor{NewColor}{Methods based on sequential \ac{mc} sampling, such as the family of \acp{pf} \cite{gordon1993novel, del1997nonlinear, liu1998sequential}, were introduced for state estimation in \acl{nl}, non-Gaussian \ac{ss} models.
To date, the \ac{kf} and its \acl{nl} variants are still widely used for online filtering in numerous \acl{rw} applications involving tracking and localization \cite{auger2013industrial}.}
%
%

The common thread among these aforementioned filters is that they are \emph{\ac{mb}} algorithms; namely, they rely on accurate 
knowledge and modeling of the underlying dynamics as a fully characterized \ac{ss} model. As such, the performance of these \ac{mb} methods critically depends on the validity of the domain knowledge and model assumptions. \ac{mb} filtering algorithms designed to cope with some level of uncertainty in the \ac{ss} models, e.g., \cite{zorzi2016robust,zorzi2017robustness,longhini2021learning}, are rarely capable of achieving the performance of \ac{mb} filtering with full domain knowledge, and rely on \textcolor{NewColor}{some} knowledge of how much their postulated model deviates from the true one. In many practical use cases the underlying dynamics of the system \textcolor{NewColor}{is} \acl{nl}, complex, and difficult to accurately characterize as a tractable \ac{ss} model, \textcolor{NewColor}{in which case} degradation in performance of \textcolor{NewColor}{the} \ac{mb} state estimators is expected.

%
Recent years have witnessed remarkable empirical success of \acp{dnn}  in real-life applications. These \ac{dd} parametric models  were shown to be able to catch the subtleties of complex processes and replace the need to explicitly characterize the domain of interest \cite{lecun2015deep, bengio2009learning}. Therefore, an alternative strategy to implement \acl{se}—without requiring explicit and accurate knowledge of the \ac{ss} model—is to learn this task from data using deep learning. \acp{dnn} such as \acp{rnn}—i.e.,  \ac{lstm} \cite{hochreiter1997long} and \acp{gru} \cite{chung2014empirical}—and attention mechanisms \cite{vaswani2017attention} have been shown to perform very well for time series related tasks mostly in intractable environments, \textcolor{NewColor}{by training these networks in an \acl{e2e}, model-agnostic manner from a large quantity of data}. Nonetheless, \acp{dnn} do not incorporate domain knowledge such as structured \ac{ss} models in a principled manner. Consequently, these \ac{dd} approaches require many trainable parameters and large data sets even for simple sequences \cite{zaheer2017latent} and lack the interpretability of \ac{mb} methods. These constraints limit the use of highly parametrized \acp{dnn} for \acl{rt} \acl{se} in applications embedded in hardware-limited mobile devices such as drones and vehicular systems.

%
The limitations of \ac{mb} Kalman filtering and \ac{dd} \acl{se} motivate a hybrid approach that exploits the best of both worlds; i.e., the soundness and low complexity of the classic \ac{kf}, and the model-agnostic nature of \acp{dnn}. Therefore, we build upon the success of our previous work in \ac{mb} \acl{dl} for signal processing and digital communication applications \cite{shlezinger2019viterbinet, shlezinger2019deepSIC, shlezinger2020learned, shlezinger2020model} to propose a hybrid \ac{mb}/\ac{dd} online recursive filter, coined \acl{kn}. In particular, we focus on \acl{rt} \acl{se} for continuous-value \ac{ss} models for which the \ac{kf} and its variants are designed. We assume that the noise statistics are unknown and the underlying \ac{ss} model is partially known or approximated from a physical model of the system dynamics. To design \acl{kn}, we identify the \ac{kg} computation of the \ac{kf} as a critical component encapsulating the dependency on noise statistics and domain knowledge, and replace it with a compact \ac{rnn} of limited complexity that is integrated into the \ac{kf} flow. The resulting system uses labeled data to learn to carry out Kalman filtering in a supervised manner.

Our main contributions are summarized as follows: 
\begin{enumerate}
\item We design \acl{kn}, which is an interpretable, low complexity, and data-efficient \ac{dnn}-aided  \acl{rt} state estimator. \acl{kn} builds upon the flow and theoretical principles of the \ac{kf},  incorporating partial domain knowledge of the underlying \ac{ss} model in its operation.  
\item By learning the \ac{kg}, \acl{kn} circumvents the dependency of the \ac{kf} on  knowledge of the underlying noise statistics, thus bypassing numerically problematic matrix inversions involved in the \ac{kf} equations and overcoming the need for tailored solutions for \acl{nl} systems; e.g., approximations to handle non-linearities as in the \ac{ekf}.
\item {We show that \acl{kn} learns to carry out \acl{kfing} from data in a manner that is invariant to the sequence length. Specifically, we present an efficient supervised training scheme that enables \acl{kn} to operate with arbitrary long trajectories while only training using short trajectories.}
\item We evaluate \acl{kn} in various \ac{ss} models. The experimental scenarios include synthetic setups, tracking the chaotic Lorenz system, and localization using the Michigan \acs{nclt} \acl{ds} \cite{carlevaris2016university}. \acl{kn} is shown to converge much faster compared \textcolor{NewColor}{with} purely \ac{dd} systems, while outperforming the \ac{mb} \ac{ekf}, \textcolor{NewColor}{\ac{ukf}, and \ac{pf},} when facing model mismatch and dominant non-linearities. 
\end{enumerate} 
The proposed \acl{kn} leverages data and partial domain knowledge to  \emph{learn the filtering operation}, rather than using data to explicitly estimate the missing \ac{ss} model parameters. Although there is a large body of work that \textcolor{NewColor}{combines} \ac{ss} models with \acp{dnn}, e.g., \cite{krishnan2015deep, karl2016deep, fraccaro2017disentangled, naesseth2018variational, archer2015black, krishnan2017structured, satorras2019combining}, these approaches are sometimes used for different \ac{ss} related tasks (e.g., smoothing, imputation);  with a different focus, e.g., incorporating high-dimensional visual observations to a \ac{kf}; or under different assumptions, as we discuss in detail below.

The rest of this paper is organized as follows: 
Section~\ref{sec:sysModel} reviews the \ac{ss} model and its associated tasks, and discusses related works. Section~\ref{sec:KNet} \textcolor{NewColor}{details the proposed  \acl{kn}}. Section~\ref{sec:Results} presents the numerical study. Section~\ref{sec:Conclusions} provides concluding remarks and future work.

Throughout the paper, we use boldface lower-case letters for vectors and boldface upper-case letters for matrices. The transpose, $\ell_2$ norm, and stochastic expectation are denoted by $\set{\cdot}^\top$,  $\norm{\cdot}$, and  $\expecteds{\cdot}$, respectively. The Gaussian distribution with mean $\gmat{\mu}$ and covariance $\gmat{\Sigma}$ is denoted by $\mathcal{N}(\gmat{\mu}, \gmat{\Sigma})$. Finally, $\greal$ and $\gint$ are the sets of real and integer numbers, respectively.

%
%
\section{System Model and Preliminaries}\label{sec:sysModel}
%
%
%
\subsection{State Space Model}\label{ssec:SSModel}
We consider dynamical systems  characterized by a \ac{ss} model in \acl{dt} \cite{bar2004estimation}. 
We focus on (possibly) \acl{nl}, Gaussian, and continuous \ac{ss} models, which for each  $t\in\gint$ are represented via 
\begin{subequations}\label{eq:NL_SS_model}
\begin{align}\label{eqn:stateEvolution}
\gvec{x}_{t}&= 
\gvec{f}\brackets{\gvec{x}_{t-1}}+\gvec{w}_{t},
&\gvec{w}_t\sim
\mathcal{N}\brackets{\gvec{0},\gvec{Q}},
\hspace{0.25cm}
&\gvec{x}_{t}\in\greal^m,\\ \label{eqn:stateObservation}
\gvec{y}_{t}&=
\gvec{h}\brackets{\gvec{x}_{t}}+\gvec{v}_{t},
&\gvec{v}_t\sim
\mathcal{N}\brackets{\gvec{0},\gvec{R}},
\hspace{0.25cm}
&\gvec{y}_{t}\in\greal^n.    
\end{align}
\end{subequations}
In \eqref{eqn:stateEvolution}, $\gvec{x}_{t}$ is the latent state vector of the system at time $t$, which evolves from the previous state $\gvec{x}_{t-1}$, by a (possibly) \acl{nl}, state-evolution function $\gvec{f}\brackets{\cdot}$ and by an \ac{awgn} $\gvec{w}_t$ with  covariance matrix $\gvec{Q}$. In \eqref{eqn:stateObservation}, $\gvec{y}_{t}$ is the vector of observations at time $t$, which is generated from the current latent state vector by a (possibly) \acl{nl} observation (emission) mapping $\gvec{h}\brackets{\cdot}$ corrupted by\ac{awgn} $\gvec{v}_t$ with covariance $\gvec{R}$. For the special case where the evolution or the observation transformations are linear,  there  exist matrices $\gvec{F},\gvec{H}$ such that
\begin{equation}\label{eqn:LinearSS}
\gvec{f}\brackets{\gvec{x}_{t-1}}={\gvec{F}}\cdot\gvec{x}_{t-1},
\quad
\gvec{h}\brackets{\gvec{x}_{t}}={\gvec{H}}\cdot\gvec{x}_{t}.   
\end{equation}
%

In practice, the state-evolution model \eqref{eqn:stateEvolution} is determined by the complex dynamics of the underlying system, while the observation model \eqref{eqn:stateObservation} is dictated by the type and quality of the observations. For instance, $\gvec{x}_t$ can determine the location, velocity, and acceleration of a vehicle, while $\gvec{y}_{t}$ are measurements obtained from several  sensors. \textcolor{NewColor}{The parameters of these models may be unknown and often require the introduction of dedicated mechanisms for their estimation in real-time \cite{yuen2016online,mu2017stable}. In some scenarios, one is likely to have access to an approximated or mismatched characterization of the underlying dynamics.}
  
%
%
%

\ac{ss} models are studied 
in the context of several different tasks; these tasks are different in their nature, and can be roughly classified into two main categories: observation \emph{approximation} and hidden state \emph{recovery}.
The first category  deals with approximating parts of the observed signal $\gvec{y}_t$. This can correspond, for example, to the prediction of future observations given past observations; the generation of missing observations in a given block via imputation; and the denoising of the observations. The second category considers the recovery of a hidden state vector $\gvec{x}_t$. This family of state recovery tasks includes offline recovery, also referred to as smoothing, where one must recover a block of hidden state vectors, given a block of observations, e.g., \cite{satorras2019combining}. The focus of this paper is \emph{filtering}; i.e.,  \emph{online} recovery of $\gmat{x}_t$  from past and current noisy observations $\{\gmat{y}_\tau\}_{\tau =1}^{t}$. For a given $\gvec{x}_0$, filtering involves the design of a mapping from $\gvec{y}_t$ to $\hat{\gvec{x}}_t$, $\forall t\in\set{1,2,\ldots,T} \triangleq \mathcal{T}$, where $T$ is the time horizon.

\subsection{Data-Aided Filtering Problem Formulation}\label{subsec:problem}
The \emph{filtering} problem is at the core of \acl{rt} tracking. Here, one must provide an instantaneous estimate of the state ${\gvec{x}}_t$ based on each incoming observation $\gvec{y}_t$ in an \emph{online} manner. 
Our main focus is on scenarios where one has \emph{partial} knowledge of the \ac{ss} model that describes the underlying dynamics. Namely, we know (or have an approximation of) the state-evolution (transition) function $\gevol$ and the state-observation (emission) function $\gobs$. For \acl{rw} applications, this knowledge is derived from our understating of the system dynamics, its physical design, and the model of the sensors. As opposed to the classical assumptions in \ac{kf}, the noise statistics $\gvec{Q}$ and $\gvec{R}$ are not known. More specifically, we assume:
\begin{itemize}
\item Knowledge of the distribution of the noise signals $\gvec{w}_t$ and $\gvec{v}_t$ is not available.
%
%
\item The functions $\gevol$ and $\gobs$ \textcolor{NewColor}{may constitute an approximation of the true underlying dynamics. Such approximations can correspond, for instance, to the  representation of continuous time dynamics in discrete time, acquisition using misaligned sensors, and other forms of  mismatches.}
\end{itemize}

\textcolor{NewColor}{
While we focus on filtering in partially known \ac{ss} models, we assume that we have access to a labeled \acl{ds} containing a sequence of observations and their corresponding \acl{gt} states. In various scenarios of interest, one can assume access to some \acl{gt} measurements in the design stage. For example, in field experiments it is possible to add extra sensors both internally or externally to collect the \acl{gt} needed for training. 
It is also possible to compute the \acl{gt} data using offline and more computationally intensive algorithms.}
%
%
%
%
Finally, the inference complexity of the learned filter  should be of the same order (and preferably smaller) as that of \ac{mb} filters, such as the \ac{ekf}.

%
%
\subsection{Related Work}\label{sec:rWork}
A key ingredient in  recursive Bayesian  filtering  is the \emph{update} operation; namely, the need to {update} the prior estimate using new observed information. For \acl{lg} \ac{ss} using the \ac{kf}, this boils down to computing the \ac{kg}. While the \ac{kf} assumes linear \ac{ss} models, many problems encountered in practice are governed by \acl{nl} dynamics, for which one should resort to approximations. 
Several extensions of the \ac{kf} were proposed to deal with non-linearities. The \ac{ekf} \cite{gruber1967approach, larson1967application} is a quasi-linear algorithm based on an analytical linearization of the \ac{ss} model. More recent \acl{nl} variations are based on numerical integration: \ac{ukf} \cite{julier1997new}, the Gauss-Hermite Quadrature \cite{arasaratnam2007discrete}, and the Cubature \ac{kf} \cite{arasaratnam2009cubature}. 
For more complex \ac{ss} models, and when the noise cannot be modeled as Gaussian, multiple variants of the \ac{pf} were proposed that are based on sequential \ac{mc} \cite{gordon1993novel, del1997nonlinear, liu1998sequential, arulampalam2002tutorial,  chopin2013smc2, martino2018distributed, andrieu2010particle, elfring2021particle}. These \ac{mc} algorithms are considered to be asymptotically exact but relatively computationally heavy when compared \textcolor{NewColor}{to} Kalman-based algorithms. These \ac{mb} algorithms require accurate knowledge of the \ac{ss} model, and their performance is typically degrades in the presence of model mismatch. 

The combination of \acl{ml} and \ac{ss} models, and specifically {Kalman}-based algorithms, is the focus of growing research attention. To frame the current work in the context of existing literature, we focus on the approaches that preserve the general structure of the \ac{ss} model. The conventional approach to deal with partially known \ac{ss} models is to impose a parametric model and then estimate its parameters. This can be achieved by {jointly} learning the parameters and state sequence using \acl{em} \cite{shumway1982approach, ghahramani1996parameter, dauwels2009expectation} and Bayesian probabilistic algorithms   \cite{yuen2016online,mu2017stable}, or by selecting from a set of \textit{a priori} known models \cite{martino2017cooperative}. When training data is available, it is commonly used to tune the missing parameters in advance, in a supervised or an unsupervised manner, as done in \cite{abbeel2005discriminative,xu2021ekfnet,barratt2020fitting}. \textcolor{NewColor}{The main drawback of these strategies is that they are restricted to an imposed parametric model on the underlying dynamics (e.g., Gaussian noises)}. 

When one can bound the uncertainty in \textcolor{NewColor}{the} \ac{ss} model in advance, an alternative approach to learning is to minimize the {worst-case} estimation error among all expected \ac{ss} models. Such robust variations were proposed for various state estimation algorithms, including Kalman variants \cite{xie1994robust,zorzi2016robust, zorzi2017robustness, longhini2021learning} and particle filters \cite{carvalho2010particle,urteaga2016sequential}. The fact that these approaches aim to design the filter to be suitable for multiple different \ac{ss} models typically results in degraded performance compared \textcolor{NewColor}{to} operating with known dynamics.

When the underlying system's dynamics are complex and only partially known or the emission model is intractable and cannot be captured in a closed form—e.g., visual observations as in a computer vision task \cite{zhou2020kfnet}—one can resort to approximations and to the use of \acp{dnn}. Variational inference \cite{kingma2013auto, rezende2014stochastic, blei2017variational} is commonly used in connection with \ac{ss} models, as in \cite{krishnan2015deep, archer2015black, karl2016deep, krishnan2017structured, fraccaro2017disentangled}, by casting the Bayesian inference task to optimization of a parameterized posterior and maximizing an objective. Such approaches cannot typically be applied directly to state recovery in \acl{rt}, as we consider here, and the learning procedure tends to be complex and prone to approximation errors. 

A common strategy when using \acp{dnn} is to encode the observations into some latent space \textcolor{NewColor}{that is assumed} to obey a simple \ac{ss} model, typically a linear Gaussian one, and track the state in the latent domain as in \cite{haarnoja2016backprop, laufer2018hybrid, zhou2020kfnet}, or to use \acp{dnn} to estimate the parameters of the \ac{ss} model as in \cite{coskun2017long, rangapuram2018deep}. Tracking in the latent space can also be extended by applying a \ac{dnn} decoder to the estimated state to return to the observations domain, while training the overall system end-to-end \cite{fraccaro2017disentangled, becker2019recurrent}. The latter allows to design trainable systems for recovering missing observations and predicting future ones by assuming that the temporal relationship can be captured as an \ac{ss} model in the latent space. This form of \ac{dnn}-aided systems is typically designed for unknown or highly complex \ac{ss} models, while we focus in this work on setups with partial domain knowledge, as detailed in Subsection~\ref{subsec:problem}. 
Another approach is to combine \acp{rnn} \cite{zheng2017state}, or variational inference \cite{salimans2015markov, naesseth2018variational} with \ac{mc} based sampling. 
Also related is the work \cite{satorras2019combining}, which used learned models in parallel with \acp{mb} algorithms operating with full knowledge of the \ac{ss} model, applying  a graph neural network in parallel to the Kalman smoother to improve its accuracy via neural augmentation. Estimation was performed by an iterative message passing over the entire time horizon. This approach is suitable for the smoothing task and is computationally intensive, and so may not be suitable for \acl{rt} filtering \cite{ni2021rtsnet}. 
%
%
\subsection{Model-Based Kalman Filtering}\label{ssec:MBKF}
Our proposed \acl{kn}, detailed in the following section, is based on the \ac{mb} \ac{kf}, which is a linear recursive estimator. In every time step $t$, the \ac{kf} produces a new estimate $\gvec{x}_t$ using only the previous estimate $\hat{\gvec{x}}_{t-1}$ as a sufficient statistic and the new observation $\gvec{y}_{t}$. As a result, the computational complexity of the \ac{kf} does not grow in time.  We first describe the original algorithm for linear \ac{ss} models, as in \eqref{eqn:LinearSS}, and then discuss how it is extended into the \ac{ekf} for  \acl{nl} \ac{ss} models. 

The \ac{kf} can be described by a two-step procedure: \textit{prediction} and \textit{update}, where in each time step $t\in \mathcal{T}$, it computes the first- and second-order statistical moments.
\begin{enumerate}
\item The first step \emph{predicts} the current \textit{a priori} statistical moments based on the previous \textit{a posteriori} estimates. Specifically, the moments of $\gvec{x}$ are computed using the knowledge of the evolution matrix $\gvec{F}$ as
\begin{subequations}\label{eqn:predict_evol}
\begin{align}\label{eqn:evol_1}
\hat{\gvec{x}}_{t\given{t-1}} &= 
\gvec{F}\cdot{\hat{\gvec{x}}_{t-1\given{t-1}}},\\\label{eqn:evol_2}
\mySigma_{t\given{t-1}} &=
{\gvec{F}}\cdot\mySigma_{t-1\given{t-1}}\cdot{\gvec{F}}^\top+\gvec{Q}
\end{align}
\end{subequations}
and the moments of the observations $\gvec{y}$ are computed based on the knowledge of the observation matrix $\gvec{H}$ as
\begin{subequations}\label{eqn:predict_obs}
\begin{align}\label{eqn:obs_1}
\hat{\gvec{y}}_{t\given{t-1}} &=
\gvec{H}\cdot\hat{\gvec{x}}_{t\given{t-1}}\\\label{eqn:obs_2}
{\gvec{S}}_{t\given{t-1}} &=
{\gvec{H}}\cdot\mySigma_{t\given{t-1}}\cdot{\gvec{H}}^\top+\gvec{R}.
\end{align}
\end{subequations}
\item In the \textit{update} step, the \textit{a posteriori} state moments are computed based on the \textit{a priori} moments as
\begin{subequations}\label{eqn:update}
\begin{align}\label{eqn:update1}
\hat{\gvec{x}}_{t\given{t}}&=
\hat{\gvec{x}}_{t\given{t-1}}+\Kgain_{t}\cdot\Delta\gvec{y}_t\\\label{eqn:update2}
{\mySigma}_{t\given{t}}&=
{\mySigma}_{t\given{t-1}}-\Kgain_{t}\cdot{\mathbf{S}}_{t\given{t-1}}\cdot\Kgain^{\top}_{t}.
\end{align}
\end{subequations}
Here, $\Kgain_{t}$ is the \ac{kg}, and it is given by
\begin{equation}\label{eq:FWGain}
\Kgain_{t}={\mySigma}_{t\given{t-1}}\cdot{\gvec{H}}^\top\cdot{\gvec{S}}^{-1}_{t\given{t-1}}.
\end{equation}
The term $\Delta\gvec{y}_t$ is the innovation; i.e., the difference between the predicted observation \textcolor{NewColor}{and} the observed value, and it is the only term that depends on the observed data
\begin{equation}\label{eqn:innov}
\Delta\gvec{y}_t=\gvec{y}_t-\hat{\gvec{y}}_{t\given{t-1}}.
\end{equation}
\end{enumerate}

The \ac{ekf} extends the \ac{kf} for \acl{nl}  $\gvec{f}\brackets{\cdot}$ and/or $\gvec{h}\brackets{\cdot}$, as in \eqref{eq:NL_SS_model}. Here, the first-order statistical moments \eqref{eqn:evol_1} and \eqref{eqn:obs_1} are replaced with
\begin{subequations}\label{eqn:predict_NL}
\begin{align}\label{eqn:evol_NL}
\hat{\gvec{x}}_{t\given{t-1}} &= 
\gvec{f}\brackets{\hat{\gvec{x}}_{t-1}},\\\label{eqn:obs_NL}
\hat{\gvec{y}}_{t\given{t-1}} &=
\gvec{h}\brackets{\hat{\gvec{x}}_{t\given{t-1}}},
\end{align}
\end{subequations}
respectively. The second-order moments, though, cannot be propagated through the non-linearity, and must thus be approximated. 
The \ac{ekf} linearizes the differentiable $\gvec{f}\brackets{\cdot}$ and $\gvec{h}\brackets{\cdot}$ in a time-dependent manner using their partial derivative matrices, also known as Jacobians, evaluated at $\hat{\gvec{x}}_{t-1\given{t-1}}$ and $\hat{\gvec{x}}_{t\given{t-1}}$. Namely,
\begin{subequations}\label{eqn:Jacob}
\begin{align}
\hat{\gvec{F}}_t&=\jacob_\gvec{f}\brackets{\hat{\gvec{x}}_{t-1\given{t-1}}}\\
\hat{\gvec{H}}_t&=\jacob_\gvec{h}\brackets{\hat{\gvec{x}}_{t\given{t-1}}},
\end{align}
\end{subequations} 
where $\hat{\gvec{F}}_t$ is plugged into \eqref{eqn:evol_2} and $\hat{\gvec{H}}_t$ is used in  \eqref{eqn:obs_2} and \eqref{eq:FWGain}. When the \ac{ss} model is linear, the \ac{ekf} coincides with the \ac{kf}, which achieves the \ac{mmse} for linear Gaussian \ac{ss} models. 

\begin{figure}  
\includegraphics[width=1\columnwidth]{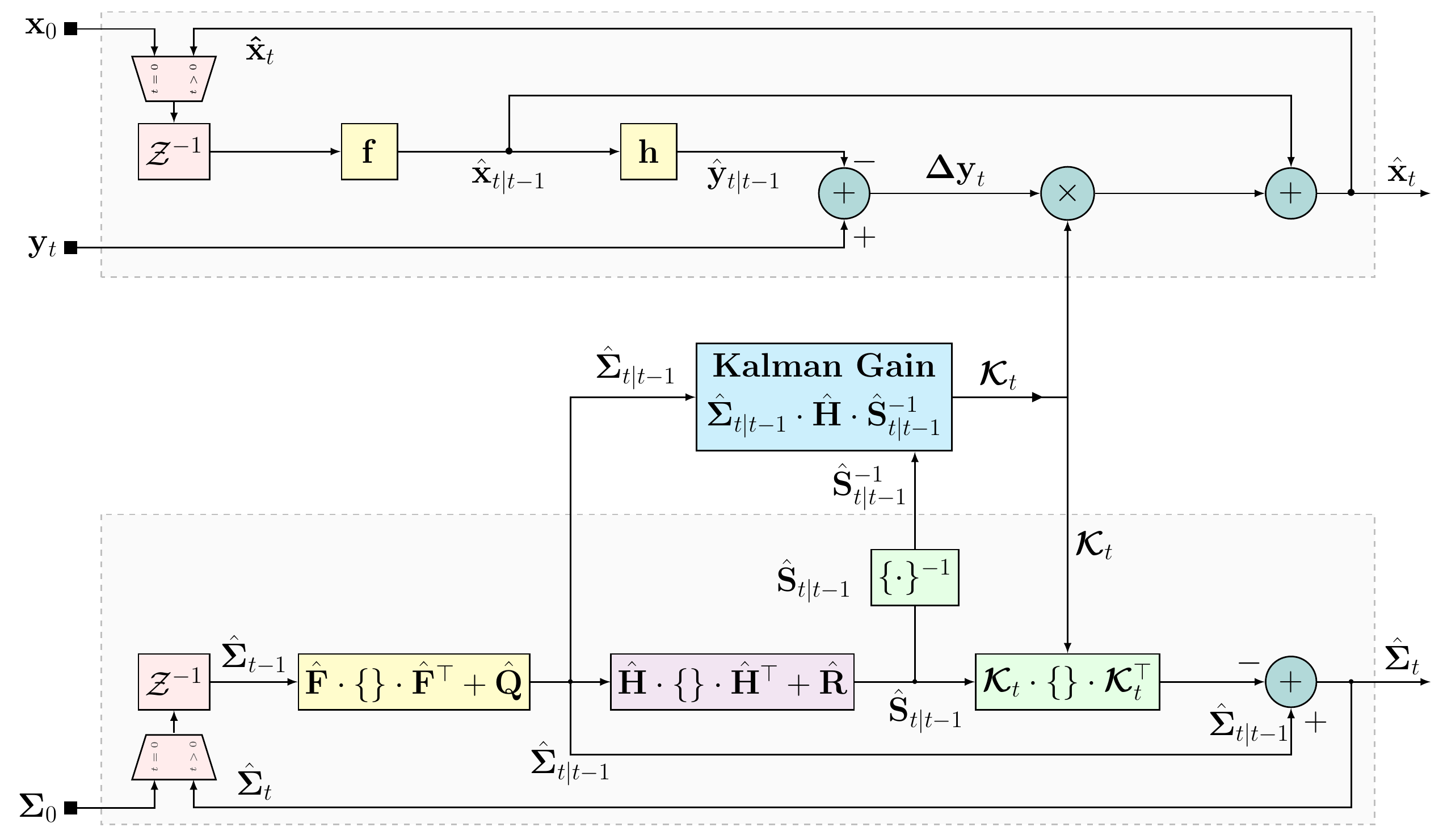}
\caption{\ac{ekf} block diagram. Here, $\mathcal{Z}^{-1}$ is the unit delay.}
\label{fig:EKF}
\end{figure}

An illustration of the \ac{ekf} is depicted in Fig.~\ref{fig:EKF}.  The resulting filter admits an efficient linear recursive structure.
However, it requires full knowledge of the underlying model and notably degrades in the presence of model mismatch. When the model is highly \acl{nl}, the local linearity approximation may not hold, and the \ac{ekf} can result in degraded performance. This motivates the augmentation of the \ac{ekf} into the deep learning-aided \acl{kn}, detailed \textcolor{NewColor}{next}. 
%
%

%
\section{KalmanNet}
\label{sec:KNet}
Here, we present \emph{\acl{kn}}; a hybrid, interpretable, data efficient architecture for \acl{rt} \acl{se} in \acl{nl} dynamical systems with partial domain knowledge. \acl{kn} combines \ac{mb} Kalman filtering with an \ac{rnn} to cope with model mismatch and non-linearities. To introduce \acl{kn}, we begin by explaining its high level operation in Subsection~\ref{subsec:Highlevel}. Then we present the features processed by its internal \ac{rnn} and the specific architectures considered for implementing and training \acl{kn} in Subsections~\ref{subsec:Features}-\ref{subsec:training}. Finally, we provide a discussion in Subsection~\ref{subsec:discussion}.
%
%
\subsection{High Level Architecture}\label{subsec:Highlevel}
\begin{figure}
\includegraphics[width=1\columnwidth]{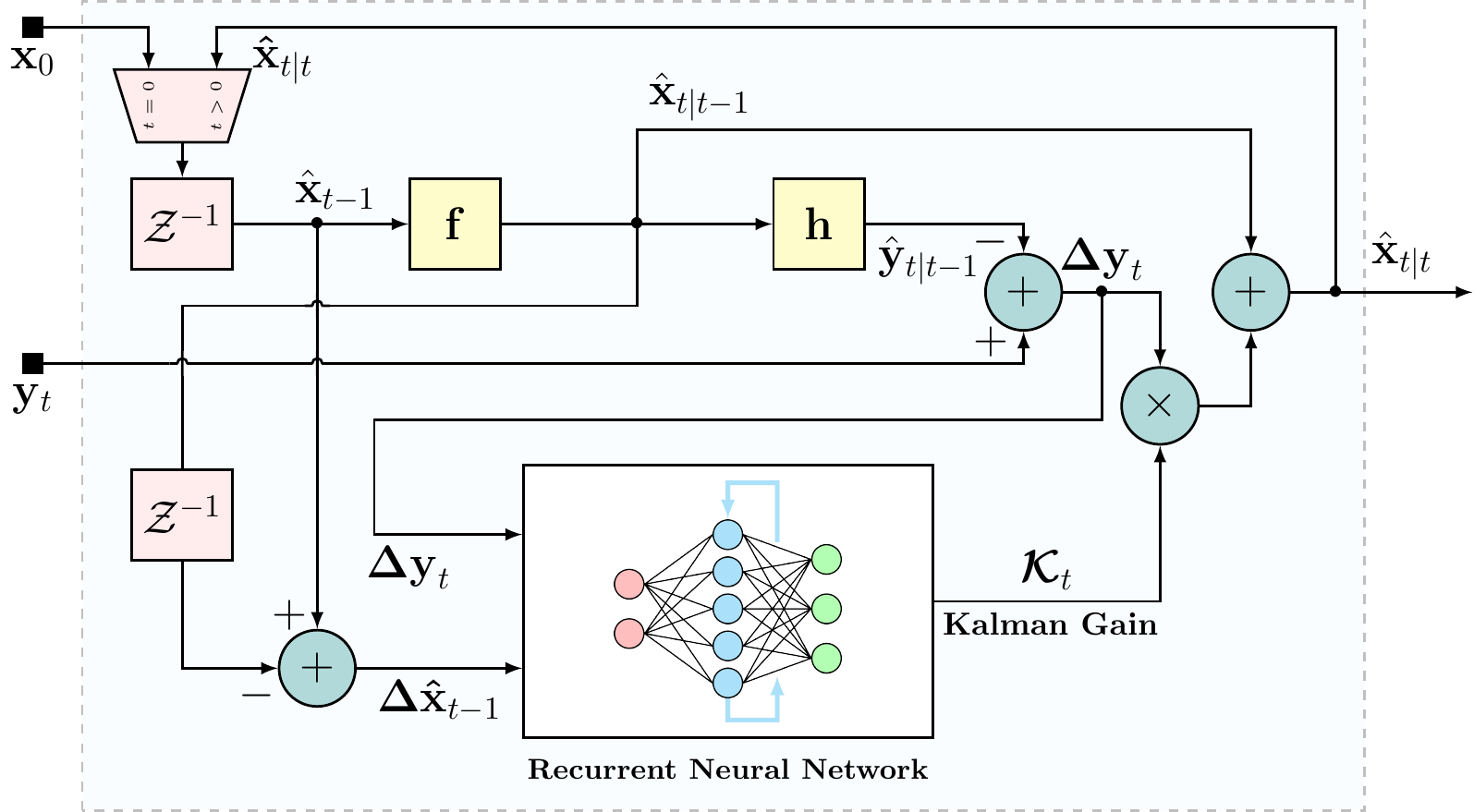}
\caption{\acl{kn} block diagram.}
\label{fig:KNet1}
\end{figure}

We formulate \acl{kn} by identifying the specific computations of the \ac{ekf} that are based on unavailable knowledge. As detailed in Subsection~\ref{subsec:problem}, the functions $\gevol$ and $\gobs$ are known (though perhaps inaccurately); yet the covariance matrices $\gvec{Q}$ and $\gvec{R}$ are unavailable. These missing statistical moments are used in \ac{mb} Kalman filtering only for computing the \ac{kg} (see Fig.~\ref{fig:EKF}). Thus, we design \acl{kn} to learn the \ac{kg} from data, and combine the learned \ac{kg} in the overall \ac{kf} flow. This high level architecture is illustrated in Fig.~\ref{fig:KNet1}.

In each time instance $t \in \mathcal{T}$, similarly to the \ac{ekf}, \acl{kn} estimates $\hat{\gvec{x}}_{t}$ in two steps; \emph{prediction} and \emph{update}. 
\begin{enumerate}
\item The \emph{prediction} step is the same as in the \ac{mb} \ac{ekf}, except that only the first-order statistical moments are predicted.  In particular, a prior estimate for the current state $\priorst$ is computed from the previous posterior $\postst{t-1}$ via  \eqref{eqn:evol_NL}. Then, a prior estimate for the current observation $\priorobs$ is computed from $\priorst$ via \eqref{eqn:obs_NL}. As opposed to its \ac{mb} counterparts, \acl{kn} does not rely on the knowledge of noise distribution and does not maintain an explicit estimate of the \acl{cov}.
\item In the \emph{update} step, \acl{kn} uses the new observation $\gvec{y}_{t}$ to  compute the current state posterior $\postst{t}$ from the previously computed prior $\priorst$ in a similar manner to the \ac{mb} \ac{kf} as in \eqref{eqn:update1}, i.e., using the innovation term $\ino{t}$ computed via \eqref{eqn:innov} and the \ac{kg} $\Kgain_t$. As opposed to the \ac{mb} \ac{ekf}, here the computation of the \ac{kg} {is not given} explicitly;  rather, it is learned from data using an \ac{rnn}, as illustrated in Fig.~\ref{fig:KNet1}. The inherent memory of \acp{rnn} allows to implicitly track the second-order statistical moments without requiring knowledge of the underlying noise statistics.
\end{enumerate}
Designing an \ac{rnn} to learn how to compute the \ac{kg} as part of an overall \ac{kf} flow requires answers to three key questions: 
\begin{enumerate}
%
%
\item From which input features (signals) will the network learn the \ac{kg}?
\item What should be the architecture of the internal \ac{rnn}?
\item How will this network be trained from data?
\end{enumerate}
In the following sections we address these questions.
%
%
%
%
\subsection{Input Features}\label{subsec:Features}
The \ac{mb} \ac{kf} and its variants compute the \ac{kg} from knowledge of the underlying statistics. To implement such computations in a learned fashion, one must provide input (features) that capture the knowledge needed to evaluate the \ac{kg} to a neural network. The dependence of $\Kgain_{t}$ on the statistics of the observations and the state process indicates that in order to track it, in every time step $t \in \mathcal{T}$, the \ac{rnn} should be provided with input containing statistical information of the observations $\gvec{y}_t$ and the state-estimate $\postst{t-1}$. Therefore, the following quantities that are related to the unknown statistical relationship of the \ac{ss} model can be used as input features to the \ac{rnn}:
\begin{enumerate}[label={\em F\arabic*}]  
\item \label{itm:obDif} The \emph{observation difference} $\Delta\tilde{\gvec{y}}_t={\gvec{y}_t-\gvec{y}_{t-1}}$.
\item \label{itm:inDif} The \emph{innovation difference} $\Delta\gvec{y}_t=\gvec{y}_t-\hat{\gvec{y}}_{t\given{t-1}}$.
\item \label{itm:FEDif} The \emph{forward evolution difference} $\Delta\tilde{\gvec{x}}_t={\hat{\gvec{x}}_{t\given{t}}-\hat{\gvec{x}}_{{t-1}\given{t-1}}}$. This quantity represents the difference between two consecutive posterior state estimates, where for time instance $t$, the available feature is $\Delta\tilde{\gvec{x}}_{t-1}$.
\item \label{itm:FUDif} The \emph{forward update difference} $\Delta\hat{\gvec{x}}_t={\hat{\gvec{x}}_{t\given{t}}-\hat{\gvec{x}}_{{t}\given{t-1}}}$, i.e., the difference between the posterior state estimate \textcolor{NewColor}{and} the prior state estimate, where again for time instance $t$ we use $\Delta\hat{\gvec{x}}_{t-1}$.
\end{enumerate}

Features \ref{itm:obDif} and \ref{itm:FEDif} encapsulate information about the state-evolution process, while features \ref{itm:inDif} and \ref{itm:FUDif} encapsulate the uncertainty of our state estimate. The difference operation removes the predictable components, and thus \textcolor{NewColor}{the time series of differences is mostly} affected by the noise statistics that we wish to learn. The \ac{rnn} described in Fig.~\ref{fig:KNet1} can use all the features, although extensive empirical evaluation suggests that the specific choice of combination of features depends on the problem at hand. Our empirical observations 
indicate that good combinations are \{\ref{itm:obDif}, \ref{itm:inDif}, \ref{itm:FUDif}\} and \{\ref{itm:obDif}, \ref{itm:FEDif}, \ref{itm:FUDif}\}. 
%
%
\subsection{Neural Network Architecture}\label{subsec:NN}
%
%
\begin{figure} 
\includegraphics[angle = -90, width=1\columnwidth]{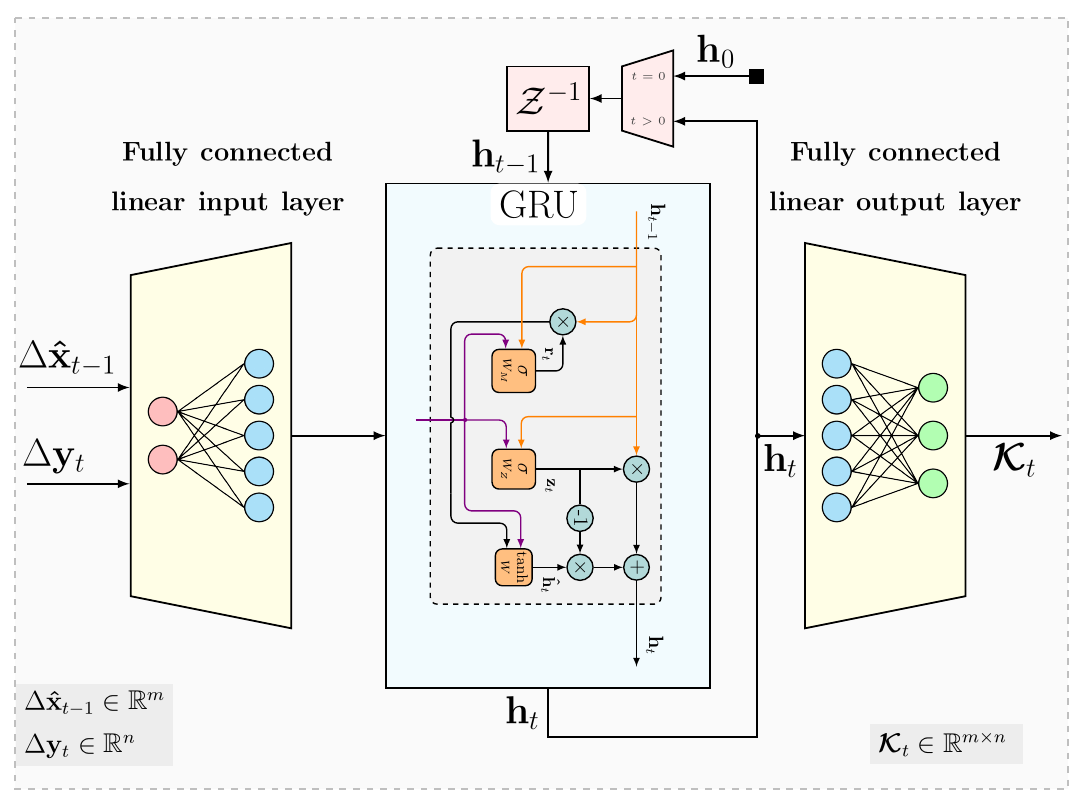}
\caption{\acl{kn} \ac{rnn} block diagram (\acl{arch1}). \textcolor{NewColor}{The architecture comprises a fully connected input layer, followed by a \ac{gru} layer (whose internal division into gates is illustrated \cite{chung2014empirical}) and an output fully connected layer. Here,} the input features are \ref{itm:inDif} and \ref{itm:FUDif}.}
\label{fig:KNet3}
\end{figure} 
The internal \ac{dnn} of \acl{kn} uses the features discussed in the previous section to compute the \ac{kg}. 
It follows from \eqref{eq:FWGain} that computing the \ac{kg} $\Kgain_{t}$ involves tracking the \acl{cov} ${\mySigma}_{t}$. The recursive nature of the \ac{kg} computation indicates that its learned module should involve an internal memory element as an \ac{rnn} to track it. 

We consider two architectures for the \ac{kg} computing \ac{rnn}. The first, illustrated in Fig.~\ref{fig:KNet3}, aims at using the internal memory of \acp{rnn} to jointly track the underlying second-order statistical moments required for computing the \ac{kg} in an implicit manner. To that aim, we use \ac{gru} cells \cite{chung2014empirical} whose hidden state is of the size of some integer product of $m^2 + n^2$, which is the joint dimensionality of the tracked moments $\hat{\gvec{\Sigma}}_{t|t-1}$ in \eqref{eqn:evol_2}, and $\hat{\gvec{S}}_t$ in \eqref{eqn:obs_2}. In particular, we first use a \ac{fc} input layer whose output is the input to the \ac{gru}. The \ac{gru} state vector $\gvec{h}_t$ is mapped into the estimated \ac{kg} $\Kgain_{t} \in \greal^{m\times n}$ using an output \ac{fc} layer with $m\cdot n$ neurons. While the illustration in Fig.~\ref{fig:KNet3} uses a single \ac{gru} layer, one can also utilize multiple layers to increase the capacity and abstractness of the network, as we do in the numerical study reported in Subsection~\ref{subsec:NCLT}. The proposed architecture does not directly design the hidden state of the \ac{gru} to correspond to the unknown second-order statistical moments that are tracked by the \ac{mb} \ac{kf}. As such, it uses a relatively large number of state variables that are expected to provide the required tracking capacity. For example, in the numerical study in Section~\ref{sec:Results} we set the dimensionality of $\gvec{h}_t$ to be $10\cdot(m^2 + n^2)$. This often results in substantial over-parameterization, as the number of \ac{gru} parameters grows quadratically with the number of state variables \cite{dey2017gate}.

\begin{figure} 
\includegraphics[width=1\columnwidth]{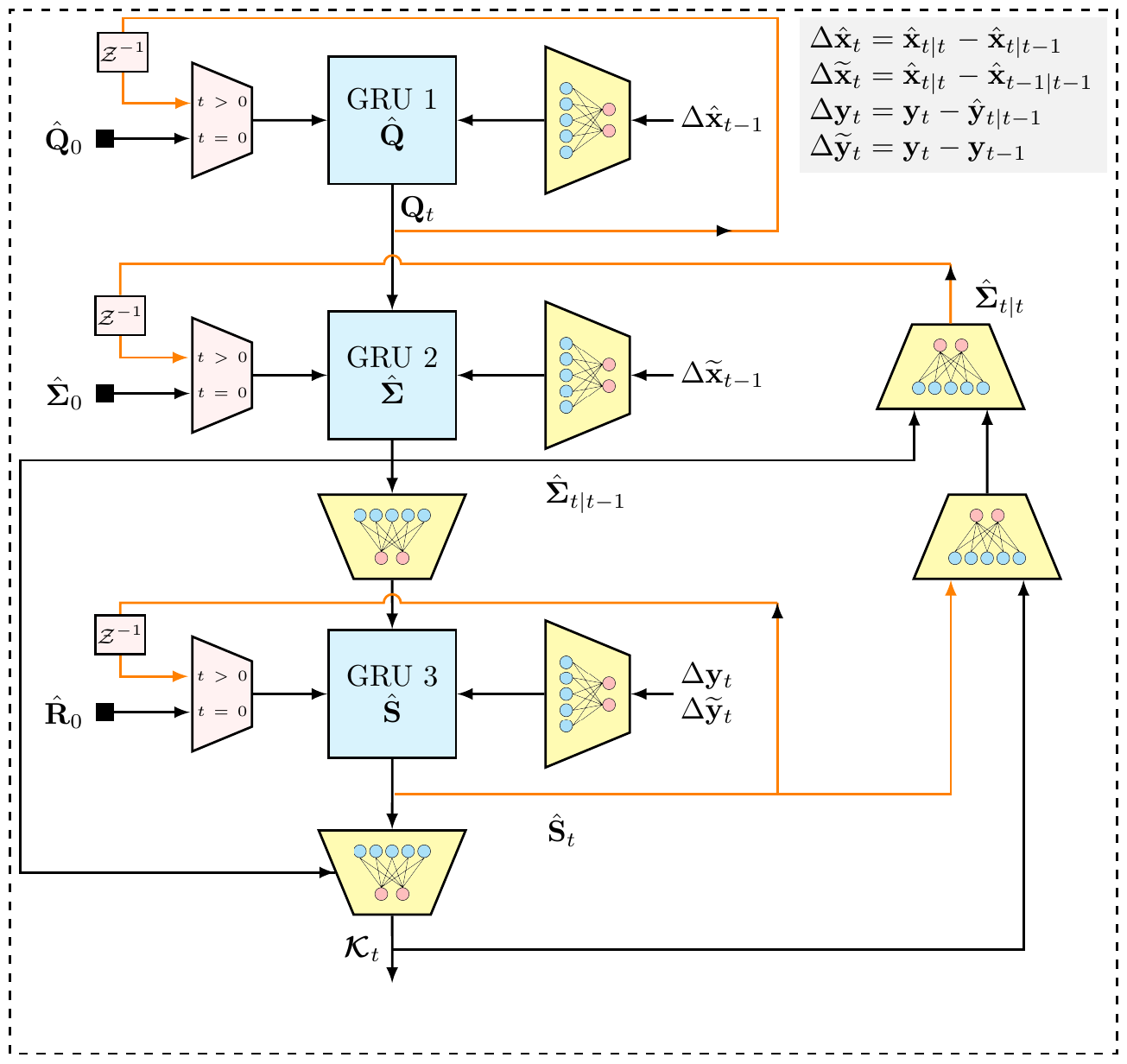}
\caption{\acl{kn} \ac{rnn} block diagram (\acl{arch2}). \textcolor{NewColor}{The input features are used to update three \acp{gru} with dedicated \ac{fc} layers, and the overall interconnection between the blocks is based on the flow of the \ac{kg} computation in the \ac{mb} \ac{kf}.}}
\label{fig:KNet2}
\end{figure}

The second architecture uses separate \ac{gru} cells for each of the tracked \acl{cov}. \textcolor{NewColor}{The division of the architecture into separate \ac{gru} cells and \ac{fc} layers and their interconnection is illustrated in Fig.~\ref{fig:KNet2}. As shown in the figure, the network} composes three \ac{gru} layers, connected in a cascade with dedicated input and output \ac{fc} layers. The first \ac{gru} layer tracks the unknown state noise covariance $\gvec{Q}$, thus tracking $m^2$ variables. Similarly, the second and third \acp{gru} track the predicted moments $\hat{\gvec{\Sigma}}_{t|t-1}$ \eqref{eqn:evol_2} and $\hat{\gvec{S}}_t$ \eqref{eqn:obs_2}, thus having $m^2$ and $n^2$ hidden state variables, respectively. The \acp{gru} are interconnected such that the learned $\gvec{Q}$ is used to compute $\hat{\gvec{\Sigma}}_{t|t-1}$, which in turn is used to obtain  $\hat{\gvec{S}}_t$, while both $\hat{\gvec{\Sigma}}_{t|t-1}$ and $\hat{\gvec{S}}_t$ are involved in producing $\Kgain_t$ \eqref{eq:FWGain}. This architecture, \textcolor{NewColor}{which is composed of a non-standard interconnection between \acp{gru} and \ac{fc} layers,} is more directly tailored towards the formulation of the \ac{ss} model and the operation of the \ac{mb} \ac{kf} compared \textcolor{NewColor}{with} the simpler first architecture. As such, it provides lesser abstraction; i.e., it is expected to be more constrained in the family of mappings it can learn compared \textcolor{NewColor}{with} the first architecture, while as a result also requiring less trainable parameters. For instance, in the numerical study reported in Subsection~\ref{subsec:Lorenz}, utilizing the first architecture requires the order of $5\cdot10^5$ trainable parameters, while the second architecture utilizes merely $2.5 \cdot 10^4$ parameters. 
%
%
%
\subsection{Training Algorithm}\label{subsec:training}
\acl{kn} is trained using the available labeled data set in a supervised manner. While we use a neural network for computing the \ac{kg} rather than for directly producing the estimate $\hat{\gvec{x}}_{t|t}$, we train \acl{kn} end-to-end. Namely, we compute the loss function $\mathcal{L}$ based on the state estimate $\postst{t}$, which is not the  output of the internal \ac{rnn}. Since this vector takes values in a continuous set $\greal^m$, we use the squared-error loss, 
\begin{equation}
\mathcal{L} = \norm{\gvec{x}_t -\hat{\gvec{x}}_{t|t}}^2    
\end{equation}
which is also used to evaluate the \ac{mb} \ac{kf}. By doing so, we build upon the ability to backpropagate the loss to the computation of the \ac{kg}. One can obtain the loss gradient with respect to the \ac{kg} from the output of \acl{kn} since
\begin{align}
\frac{\partial \mathcal{L}}{\partial\Kgain_t} &=\frac{\partial \norm{\Kgain_t \ino{t} -\dx{t}}^2}{\partial\Kgain_t } \notag \\
&=2 \cdot \brackets{\Kgain_t\cdot \ino{t} - \dx{t}}\cdot\ino{t}^\top,
\label{eqn:gradient1}
\end{align}
where $\dx{t}\triangleq \gvec{x}_t -\hat{\gvec{x}}_{t|t-1}$. The gradient computation in \eqref{eqn:gradient1}  indicates that one can learn the computation of the \ac{kg} by training \acl{kn} end-to-end using the squared-error loss. \textcolor{NewColor}{In particular, this allows to train the overall filtering system without having to externally provide ground truth values of the \ac{kg} for training purposes.}

The data set used for training comprises $N$ trajectories that can be of varying lengths. Namely, by letting $T_i$ be the length of the $i$th training trajectory, the data set is given by $\mathcal{D} = \set{(\gvec{Y}_i, \gvec{X}_{i})}_{1}^N$, where 
\begin{align}
\gvec{Y}_i&=\big[\gvec{y}_1^{\brackets{i}},\ldots,\gvec{y}_{T_i}^{\brackets{i}}\big],  \quad 
\gvec{X}_i&=\big[\gvec{x}_0^{\brackets{i}},\gvec{x}_1^{\brackets{i}},\ldots,\gvec{x}_{T_i}^{\brackets{i}}\big].
\end{align}
By letting $\NNParam$ denote the trainable parameters of the \ac{rnn}, and  $\gamma$ be a regularization coefficient, we then construct an $\ell_2$ regularized \ac{mse} loss measure
\begin{equation}
\ell_i\brackets{\NNParam}=
\frac{1}{T_i}\sum_{t=1}^{T_i}\norm{
\hat{\gvec{x}}_t\brackets{\gvec{y}^{\brackets{i}}_t;\NNParam}\!-\! \gvec{x}^{\brackets{i}}_t}^2+
\gamma\cdot\norm{\NNParam}^2.
\label{eqn:loss2}
\end{equation}
 To optimize $\NNParam$, we use a variant of mini-batch \acl{sgd} in which for every {batch} indexed by $k$, we choose $M < N$  trajectories indexed by $i_1^k, \ldots, i_M^k$, computing the  mini-batch loss as
\begin{equation}
\mathcal{L}_k\brackets{\NNParam}=
\frac{1}{M}\sum_{j=1}^M\ell_{i_j^k}\brackets{\NNParam}.
\label{eqn:Loss}
\end{equation}
%

Since \acl{kn} is a recursive architecture with both an external recurrence and an internal \ac{rnn}, we use the \ac{bptt} algorithm \cite{werbos1990backpropagation} to train it. Specifically, we unfold \acl{kn} across time with shared network parameters, and then compute a forward and backward gradient estimation pass through the network. We consider three different variations of applying the \ac{bptt} algorithm for training \acl{kn}:
%
%
\begin{enumerate}[label={\em V\arabic*}]
\item \label{alg3} Direct application of \ac{bptt}, where for each training iteration the gradients are computed over the entire trajectory. 
\item \label{alg2} An application of the truncated \ac{bptt} algorithm \cite{sutskever2013training}. Here, given a \acl{ds} of long trajectories (e.g., $T=3000$ time steps), each long trajectory is divided into multiple short trajectories (e.g., $T=100$ time steps), which are shuffled and used during training. 
\item \label{alg1} An alternative application of truncated \ac{bptt}, where we truncate each trajectory to a fixed (and relatively short) length, and train using these short trajectories. 
\end{enumerate}

Overall, directly applying \ac{bptt} via \ref{alg3} may be computationally expensive and unstable. Therefore, a favored approach is to first use the truncated \ac{bptt} as in \ref{alg2} as a warm-up phase (train first on short trajectories) in order to stabilize its learning process, after which \acl{kn} is tuned using \ref{alg3}. The procedure in \ref{alg1} is most suitable for systems that are known to be likely to quickly converge to a steady state (e.g., linear \ac{ss} models). In our numerical study, reported in Section~\ref{sec:Results}, we utilize all three approaches.
%
%
\subsection{Discussion}\label{subsec:discussion}
\acl{kn} is designed to operate in a hybrid \ac{dd}/\ac{mb} manner, combining \acl{dl} \textcolor{NewColor}{with} the classical \textcolor{NewColor}{\ac{ekf}} procedure. By identifying the specific noise-model-dependent computations of the \ac{ekf} and replacing them with a dedicated \ac{rnn} integrated in the \textcolor{NewColor}{\ac{ekf}} flow, \acl{kn} benefits from the individual strengths of both \ac{dd} and \ac{mb} approaches. \textcolor{NewColor}{The augmentation of the \ac{ekf} with dedicated deep learning modules results in several core differences between \acl{kn} and its \ac{mb} counterpart. Unlike the \ac{mb} \ac{ekf}, \acl{kn} does not attempt to linearize the \ac{ss} model, and does not impose a statistical model on the noise signals. In addition, \acl{kn} filters in a non-linear manner, as its \ac{kg} matrix depends on the input $\gvec{y}_t$. Due to these differences,} compared to \ac{mb} Kalman filtering, \acl{kn} is more robust to model mismatch and can infer more efficiently, as demonstrated in Section \ref{sec:Results}. In particular, the \ac{mb} \ac{ekf} is sensitive to inaccuracies in the underlying \ac{ss} model, e.g., in $\gevol$ and $\gobs$, while \acl{kn} can overcome such uncertainty by learning an alternative \ac{kg} that yields accurate estimation. 

Furthermore, \acl{kn} is  derived for \ac{ss} models when noise statistics are not specified explicitly. \textcolor{NewColor}{A \ac{mb} approach to tackle this without relying on data employs the \acl{rkf} \cite{zorzi2016robust,zorzi2017robustness,longhini2021learning}, which designs the filter to minimize the maximal \ac{mse} within some range of assumed \ac{ss} models, at the cost of performance loss, compared to knowing the true model. When one has access to data,} the direct strategy to implement the \ac{ekf} in such setups is to use the data to estimate $\gvec{Q}$ and $\gvec{R}$, either directly from the data or by backpropagating through the operation of the \ac{ekf} as in \cite{xu2021ekfnet}, and utilize these estimates to compute the \ac{kg}. As covariance estimation can be a challenging task when dealing with high-dimensional signals, \acl{kn} bypasses this need by directly learning the \ac{kg}, \textcolor{NewColor}{and by doing so approaches the \ac{mse} of \ac{mb} Kalman filtering with full knowledge of the \ac{ss} model, as demonstrated in Section~\ref{sec:Results}}. Finally, the computation complexity
for each time step $t \in \mathcal{T}$ is also linear in the \ac{rnn} dimensions and does not involve matrix inversion. This implies that \acl{kn} is a good candidate to apply for high dimensional \ac{ss} \textcolor{NewColor}{models} and on computationally limited devices.

Compared \textcolor{NewColor}{to} purely \ac{dd} state estimation, \acl{kn} benefits from its model awareness and the fact that its operation follows the flow of  \ac{mb} Kalman filtering rather than being utilized as a black box. As numerically observed in Section~\ref{sec:Results}, \acl{kn} achieves improved \ac{mse} compared \textcolor{NewColor}{to} utilizing \acp{rnn} for end-to-end state estimation, and also approaches the \ac{mmse} performance achieved by the \ac{mb} \ac{kf} in linear Gaussian \ac{ss} models. 
Furthermore, the fact that \acl{kn} preserves the flow of the \ac{ekf} implies that the intermediate features exchanged between its modules have a specific operation meaning, providing interpretability that is often scarce in \acl{e2e}, \acl{dl} systems. Finally, the fact that \acl{kn} learns to compute the \ac{kg} indicates the possibility of providing not only estimates of the state $\gvec{x}_t$, but also a measure of confidence in this estimate, as the \ac{kg} can be related to the covariance of the estimate, as initially explored in \cite{klein2021uncertainty}. 

These combined gains of \acl{kn} over purely \ac{mb} and \ac{dd} approaches were recently observed in \cite{LopezICAS21}, which utilized an early version of \acl{kn}  for \acl{rt} velocity estimation in an autonomous racing car. In such a setup, a \acl{nl}, \ac{mb} mixed \ac{kf} was traditionally used, and suffered from performance degradation due to inherent mismatches in the formulation of the \ac{ss} model describing the problem. Nonetheless, previously proposed \ac{dd} techniques relying on \acp{rnn} for \acl{e2e} state estimation were not operable in the desired frequencies on the hardware limited vehicle control unit. It was shown in \cite{LopezICAS21} that the application of \acl{kn} allowed to achieve improved \acl{rt} velocity tracking compared \textcolor{NewColor}{to} \ac{mb} techniques while being deployed on the control unit of the vehicle.

Our design of \acl{kn}  gives rise to many interesting future extensions. Since we focus here on \ac{ss} models where the mappings $\gevol$ and $\gobs$ are known up to some approximation errors, a natural extension of \acl{kn} is \textcolor{NewColor}{to use the data to pre-estimate them, as demonstrated briefly in the numerical study. Another alternative to cope with these approximation errors is to utilize dedicated neural networks to learn these mappings while training the entire model in an \acl{e2e} fashion.} Doing so is expected to allow \acl{kn} to be utilized in scenarios with analytically intractable \ac{ss} models, as often arises when tracking based on unstructured observations, e.g., visual observations as in \cite{zhou2020kfnet}.

While we train \acl{kn} in a supervised manner using labeled data, the fact that it preserves the operation of the \ac{mb} \ac{ekf} that produces a prediction of the next observation $\hat{\gvec{y}}_{t|t-1}$ for each time instance indicates the possibility of using this intermediate feature for \emph{unsupervised} training. One can thus envision \acl{kn} being trained offline in a supervised manner, while tracking variations in the underlying \ac{ss} model at run-time by online self supervision, following a similar rationale to that used in \cite{shlezinger2019viterbinet,shlezinger2019deepSIC} for deep symbol detection in time-varying communication channels. 

\textcolor{NewColor}{Finally, we note that while we focus here on filtering tasks, \ac{ss} models are used to represent additional related problems such as smoothing and prediction, as discussed in Subsection~\ref{ssec:SSModel}. The fact that \acl{kn} does not explicitly estimate the \ac{ss} model implies that it cannot simply substitute these parameters into an alternative algorithm capable of carrying out tasks other than filtering.  Nonetheless, one can still design \ac{dnn}-aided algorithms for these tasks operating with partially known \ac{ss} models as extensions of \acl{kn}, in the same manner as many \ac{mb} algorithms build upon the \ac{kf}. For instance, as the \ac{mb} \ac{kf} constitutes the first part of the Rauch-Tung-Striebel smoother \cite{rauch1965maximum}, one can extend \acl{kn} to implement high-performance smoothing in partially known \ac{ss} models, as we have recently began investigating  in \cite{ni2021rtsnet}.  Nonetheless, we leave the exploration of extensions of \acl{kn} to alternative tasks associated with \ac{ss} models for future work. }  
%
%
\section{Experiments and Results}
\label{sec:Results}
In this section we present an extensive numerical study of \acl{kn}\footnote{The source code used in our numerical study along with the complete set of hyperparameters used in each numerical evaluation can be found online at \url{https://github.com/KalmanNet/KalmanNet_TSP}.}, evaluating its performance in multiple setups 
and comparing it to various benchmark algorithms:
\begin{enumerate}[label=(\alph*)]
\item In our first experimental study, we consider multiple \emph{linear} \ac{ss} models, and compare \acl{kn} to the \ac{mb} \ac{kf} which is known to minimize the \ac{mse} in such a setup. We also confirm our design and architectural choices by comparing \acl{kn} with alternative \ac{rnn} based \acl{e2e} state estimators.
\item We next consider two \emph{\acl{nl}} \ac{ss} models, a sinusoidal model, and the chaotic \acl{la}. We compare \acl{kn} with the common \acl{nl} \ac{mb} benchmarks; namely, the \ac{ekf}, \ac{ukf}, and \ac{pf}. 
\item In our last study we consider a \emph{localization} use case \textcolor{NewColor}{based on} the Michigan \acs{nclt} \acl{ds} \cite{carlevaris2016university}. \textcolor{NewColor}{Here, we compare \acl{kn} with \ac{mb} \ac{kf} that assumes a linear \textit{Wiener} \emph{kinematic} model \cite{bar2004estimation} and with a \emph{vanilla} \ac{rnn} based \acl{e2e} state estimator, and demonstrate the ability of \acl{kn} to track \acl{rw} dynamics that was not synthetically generated from an underlying \ac{ss} model.}
\end{enumerate}
%
%
\subsection{Experimental Setting}
Throughout the numerical study and unless stated otherwise, in the experiments involving synthetic data, the \ac{ss} model is generated using diagonal noise covariance matrices; i.e., 
\begin{equation}
\label{eqn:nuDef}
\gvec{Q}=\mathrm{q}^2\cdot\gvec{I},
\quad
\gvec{R}=\mathrm{r}^2\cdot\gvec{I},
\quad
\nu\triangleq\frac{\mathrm{q}^2}{\mathrm{r}^2}. 
\end{equation}
\textcolor{NewColor}{
By \eqref{eqn:nuDef}, setting $\nu$ to be $0$ dB implies that both the state noise and the observation noise have the same variance.}
For consistency, we use the term \emph{full information} for cases where the  \ac{ss} model available to  \acl{kn} and its \ac{mb} counterparts accurately represents the underlying dynamics. More specifically,  \acl{kn} operates with {full} knowledge of $\gevol$ and $\gobs$, and without access to the noise covariance matrices, while its \ac{mb} counterparts operate with an accurate knowledge of $\gvec{Q}$ and $\gvec{R}$.
The term \emph{partial information} refers to the case where \acl{kn} and its \ac{mb} counterparts operate with some level of model {mismatch}, where the \ac{ss} model design parameters do not represent the underlying dynamics accurately (i.e., are not equal to the \ac{ss} parameters from which the data was generated). 
\textcolor{NewColor}{Unless stated otherwise, the metric used to evaluate the performance is the \ac{mse} on a $\dB$ scale.} In the figures we depict the \ac{mse} in $\dB$ versus the inverse observation noise level, i.e., $\frac{1}{\gscal{r} ^2}$, also on a $\dB$ scale. \textcolor{NewColor}{In some of our experiments, we evaluate both the \ac{mse} and its standard deviation, where we denote these measures by $\hat{\mu}$ and $\hat{\sigma}$, respectively.}
%
%
\vspace{0.1cm}

\subsubsection{\acl{kn} Setting}
In Section~\ref{sec:KNet} we present several architectures and training mechanisms that can be used when implementing \acl{kn}. In our experimental study we consider three different configurations of \acl{kn}:
\begin{enumerate}[label={\em C\arabic*}]
\item \label{cfg1} \acl{kn} \acl{arch1} with input features \{\ref{itm:inDif}, \ref{itm:FUDif}\} and with training algorithm \ref{alg1}.
\item \label{cfg1a} \acl{kn} \acl{arch1} with input features \{\ref{itm:inDif}, \ref{itm:FUDif}\} and with training algorithm \ref{alg3}.
\item \label{cfg2} \acl{kn} \acl{arch1} with input features \{\ref{itm:obDif}, \ref{itm:FEDif}, \ref{itm:FUDif}\} and with training algorithm \ref{alg2}. 
\item \label{cfg3} \acl{kn} \acl{arch2} with all input features and with training algorithm \ref{alg3}. 
\end{enumerate}
In all our experiments \acl{kn} was trained using the Adam optimizer \cite{kingma2014adam}. 
%
%
\vspace{0.1cm}

\subsubsection{Model-Based Filters}\label{ssec:mb_setup}
In the following experimental study we compare \acl{kn} with several \ac{mb} filters. For the \ac{ukf} we used the software package \cite{FilterPy}, while the \ac{pf} is implemented based on \cite{pyParticleEst} using 100 particles and without parallelization. During our numerical study, when model uncertainty was introduced, we optimized the performance of the \ac{mb} algorithms by carefully tuning  the covariance matrices, usually via a grid search. For long trajectories (e.g., $T>1500$) it was sometimes necessary to tune these matrices, even in the case of full information, to compensate for inaccurate uncertainty propagation due to \acl{nl} approximations and to avoid divergence. 
%
%
\subsection{Linear State Space Model}\label{subsec:SynLinear}
Our first experimental study compares \acl{kn} to the \ac{mb} \ac{kf} for different forms of synthetically generated {linear} system dynamics. Unless stated otherwise, here $\gvec{F}$ takes the 
controllable canonical form. 
%
%
%
\vspace{0.1cm}

\subsubsection{Full Information}
\begin{figure*}
\begin{center}
%
%
\begin{subfigure}[pt]{0.99\columnwidth}
\includegraphics[width=1\columnwidth]{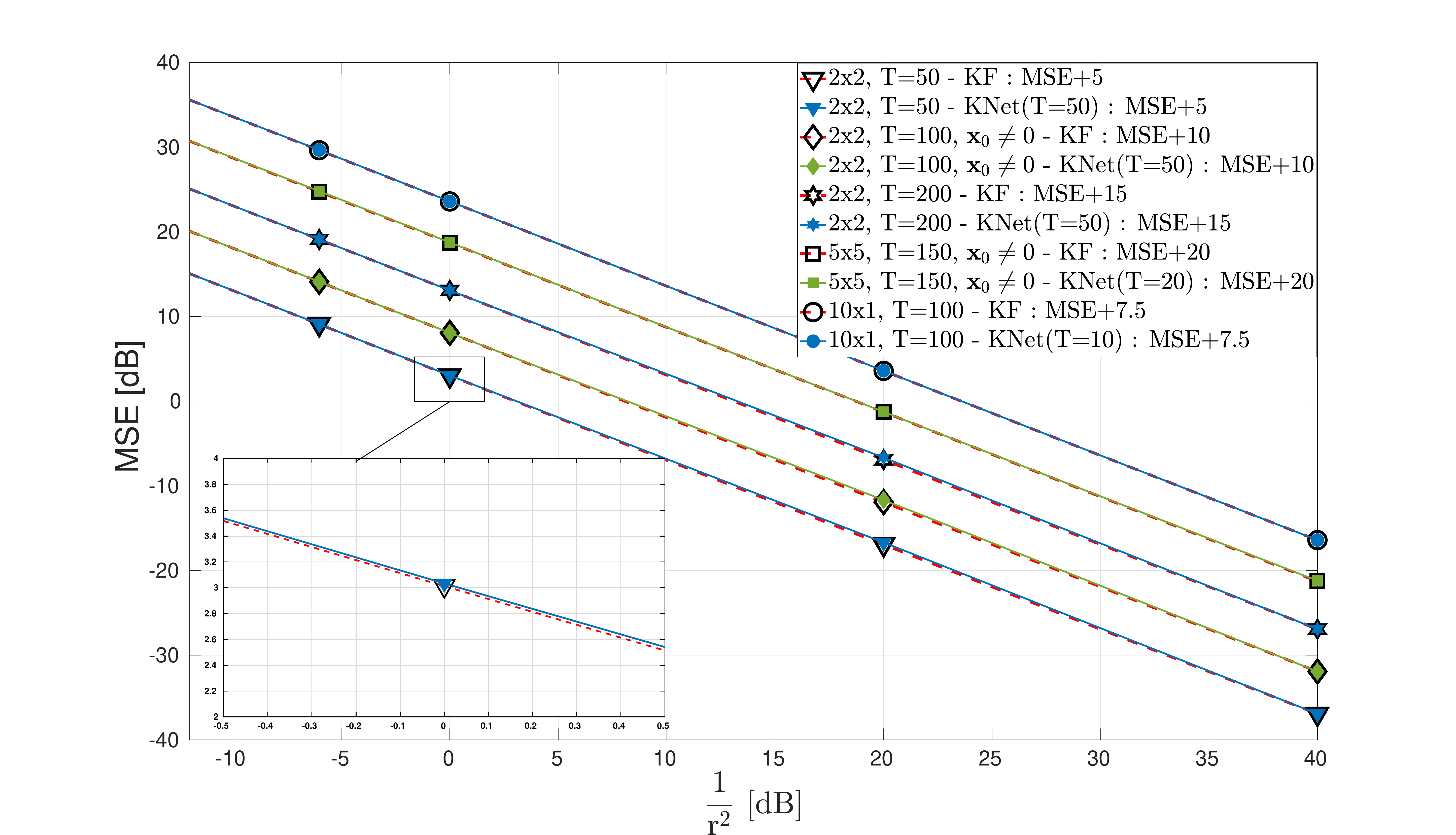}
\caption{\acl{kn} converges to \ac{mmse}.}
\label{fig:Lin_full} 
\end{subfigure}
%
%
\begin{subfigure}[pt]{0.99\columnwidth}
\includegraphics[width=1\columnwidth]{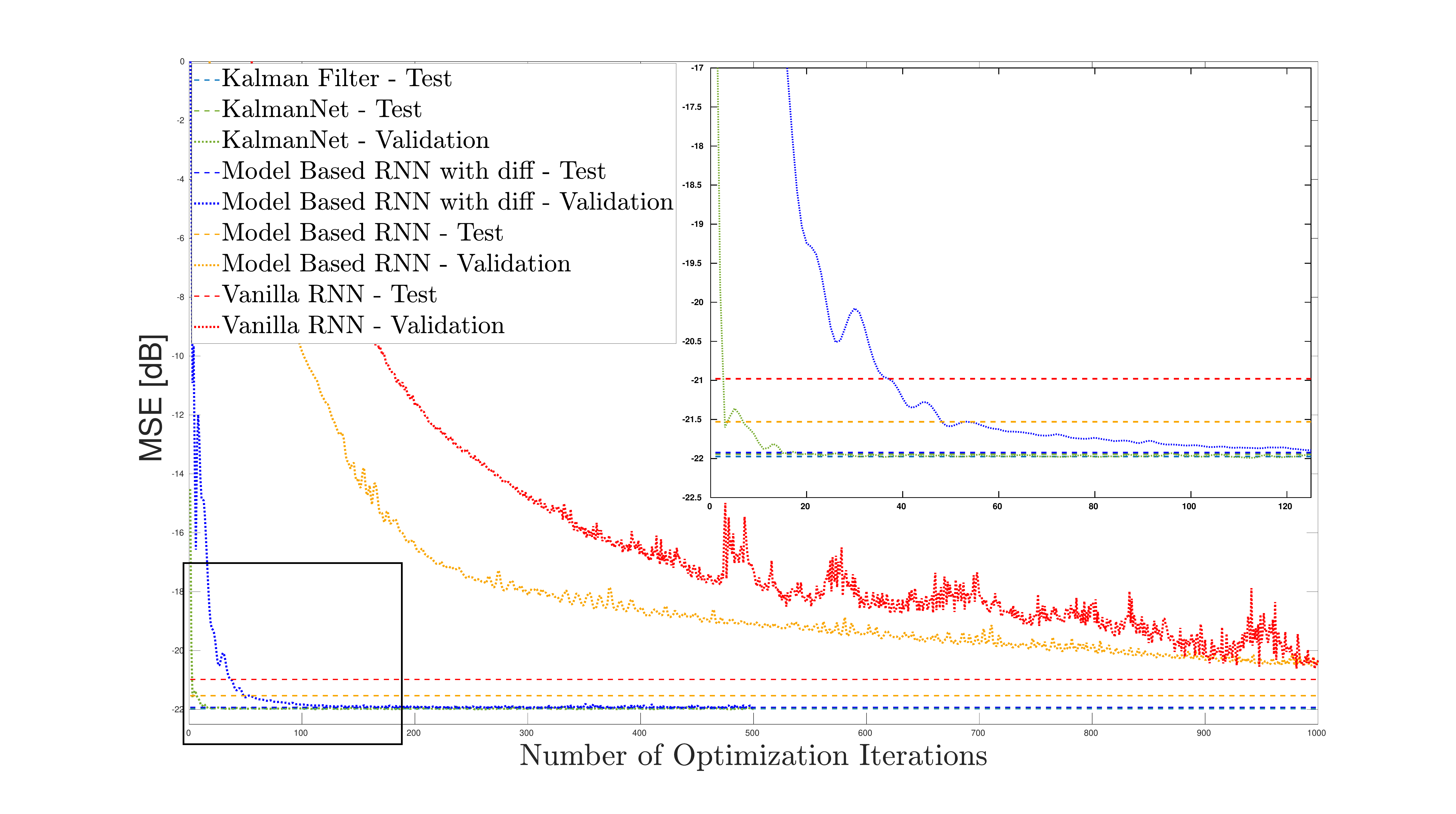}
\caption{Learning curves for \ac{dd} \acl{se}.}
\label{fig:compare}
\end{subfigure}
\caption{Linear \ac{ss} model with full information.}
\end{center}
\figSpace
\vspace{-0.2cm}
\end{figure*}

We start by comparing \acl{kn} of setting \ref{cfg1} to the \ac{mb} \ac{kf} for the case of full information, where the latter is known to minimize the \ac{mse}. Here, we set $\gvec{H}$ to take the inverse canonical form, and $\nu=0\dB$. To demonstrate the applicability of \acl{kn} to \textcolor{NewColor}{various} linear systems, we experimented with systems of different dimensions; namely, $m\times n\in\set{2\times2, 5\times5, 10 \times 1}$, and with trajectories of different lengths; namely, $T\in \set{50, 100, 150, 200}$. In Fig.~\ref{fig:Lin_full} we can clearly observe that \acl{kn} achieves the \ac{mmse} of the \ac{mb} \ac{kf}. Moreover, to further evaluate the gains of the hybrid architecture of \acl{kn}, we check that its learning is transferable. Namely, in some of the experiments, we test \acl{kn} on longer trajectories then those it was trained on, and with different initial conditions. The fact that \acl{kn} achieves the \ac{mmse} lower bound also for these cases indicates that it indeed learns to implement \acl{kfing}, and it is not tailored to the trajectories presented during training, with dependency only on the \ac{ss} model.
%
%
%
%

\subsubsection{Neural Model Selection}\label{subsec:modelSelect}
Next, we evaluate and confirm our design and architectural choices by considering a $2\times 2$ setup (similar to the previous one), and by comparing \acl{kn} with setting \ref{cfg1} to two \ac{rnn} based architectures of similar capacity applied for end-to-end state estimation:
\begin{itemize}
\item {\em Vanilla \ac{rnn}} directly maps the observed $\gvec{y}_t$ to an estimate of the state $\hat{\gvec{x}}_t$.
\item {\em \ac{mb} \ac{rnn}} imitates the Kalman filtering operation by first recovering $\hat{\gvec{x}}_{t\given{t-1}}$ using domain knowledge, i.e., via \eqref{eqn:evol_1}, and then uses the \ac{rnn} to estimate an increment  $\Delta\hat{\gvec{x}}_t$ from the prior to posterior.
\end{itemize} 
All \acp{rnn} utilize the same architecture as in \acl{kn} with a single \ac{gru} layer and the same learning hyperparameters. In this experiment we test the trained models on trajectories with the same length as \textcolor{NewColor}{they were} trained on, namely $T=20$. We can clearly observe how each of the key design considerations of \acl{kn} affect the learning curves depicted in Fig.~\ref{fig:compare}: 
\begin{itemize}
\item The incorporation of the known \ac{ss} model allows the \ac{mb} \ac{rnn} to outperform the vanilla \ac{rnn}, although both converge slowly and fail to achieve the \ac{mmse}. 
\item Using the \textcolor{NewColor}{sequences of differences} as input notably improves the convergence rate of the \ac{mb} \ac{rnn}, indicating the benefits of using the \textcolor{NewColor}{differences as features}, as discussed in Subsection~\ref{subsec:Features}. 
\item Learning is further improved by using the \ac{rnn} for recovering the \ac{kg} as part of the \ac{kf} flow, as done by \acl{kn}, rather than for directly estimating $\gvec{x}_t$. 
\end{itemize}

To further evaluate the gains of \acl{kn} over end-to-end \acp{rnn}, we compare the pre-trained models using trajectories with different initial conditions and a longer time horizon ($T=200$) than the one on which they were trained ($T=20$). The results, summarized in Table~\ref{tbl:TestMSE}, show that \acl{kn} maintains achieving the \ac{mmse}, as already observed in Fig.~\ref{fig:Lin_full}. The \ac{mb} \ac{rnn} and vanilla \ac{rnn} are more than $50\dB$ from the \ac{mmse}, implying that their learning is not transferable and that they do not learn to implement \acl{kfing}. However, when provided with the difference features as we proposed in Subsection~\ref{subsec:Features}, the \ac{dd} systems are shown to be applicable in longer trajectories, with \acl{kn} achieving \ac{mse} within a minor gap of that achieved by the \ac{mb} \ac{kf}. The results of this study validate the  considerations used in designing \acl{kn} for the \ac{dd} filtering problem discussed in Subsection~\ref{subsec:problem}. 
%
%
\begin{table}
\caption{Test \ac{mse} in [dB] when trained using  $T=20$.}
\vspace{-0.2cm}
\begin{center}
{\scriptsize
\begin{tabular}{|c|c|c|c|c| c| }
\hline
Test $T$& Vanilla \ac{rnn} & \ac{mb} \ac{rnn} & \ac{mb} \ac{rnn}, diff. & \acl{kn} & \ac{kf} \\  
\hline
$20$ & -20.98 & -21.53 & -21.92 & -21.92 & {\bf -21.97}\\  
\hline
$200$ &  58.14 &  36.8 & -21.88&  -21.90 & {\bf -21.91}\\
\hline
\end{tabular}
\label{tbl:TestMSE}
}
\end{center}
\figSpace
\end{table} 
\subsubsection{Partial Information}
To conclude our study on linear models, we next evaluate the robustness of \acl{kn} to model mismatch as a result of partial model information. We simulate a $2\times 2$ \ac{ss} model with mismatches in either the state-evolution model ($\gvec{F}$) or in the state-observation model ($\gvec{H}$). 

\begin{figure*}
\begin{center}
%
%
\begin{subfigure}[pt]{0.99\columnwidth}
\includegraphics[width=1\columnwidth]{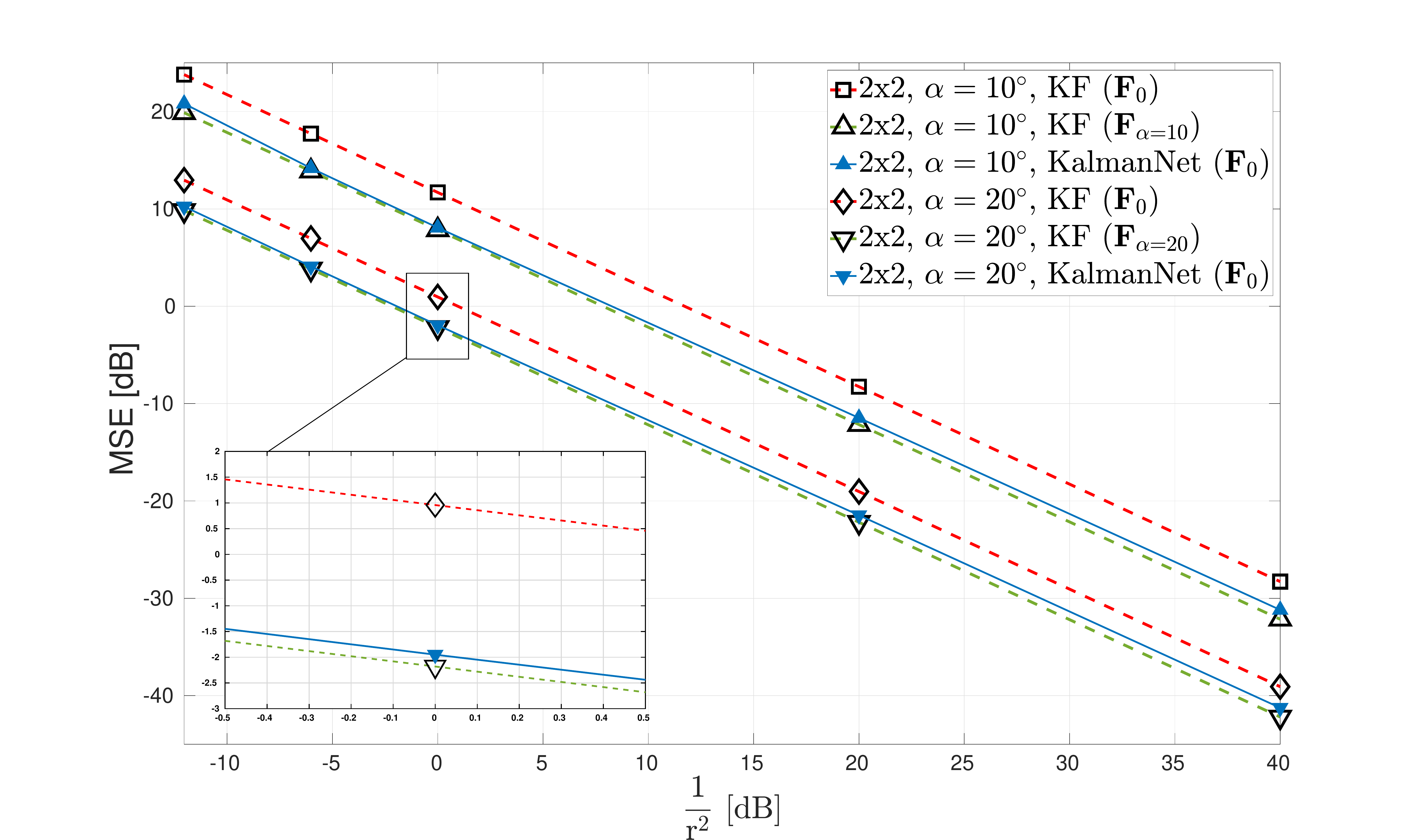}
\caption{State-evolution mismatch.} 
\label{fig:Lin_evol_mismatch}
\end{subfigure}
%
%
\begin{subfigure}[pt]{0.99\columnwidth}
\includegraphics[width=1\columnwidth]{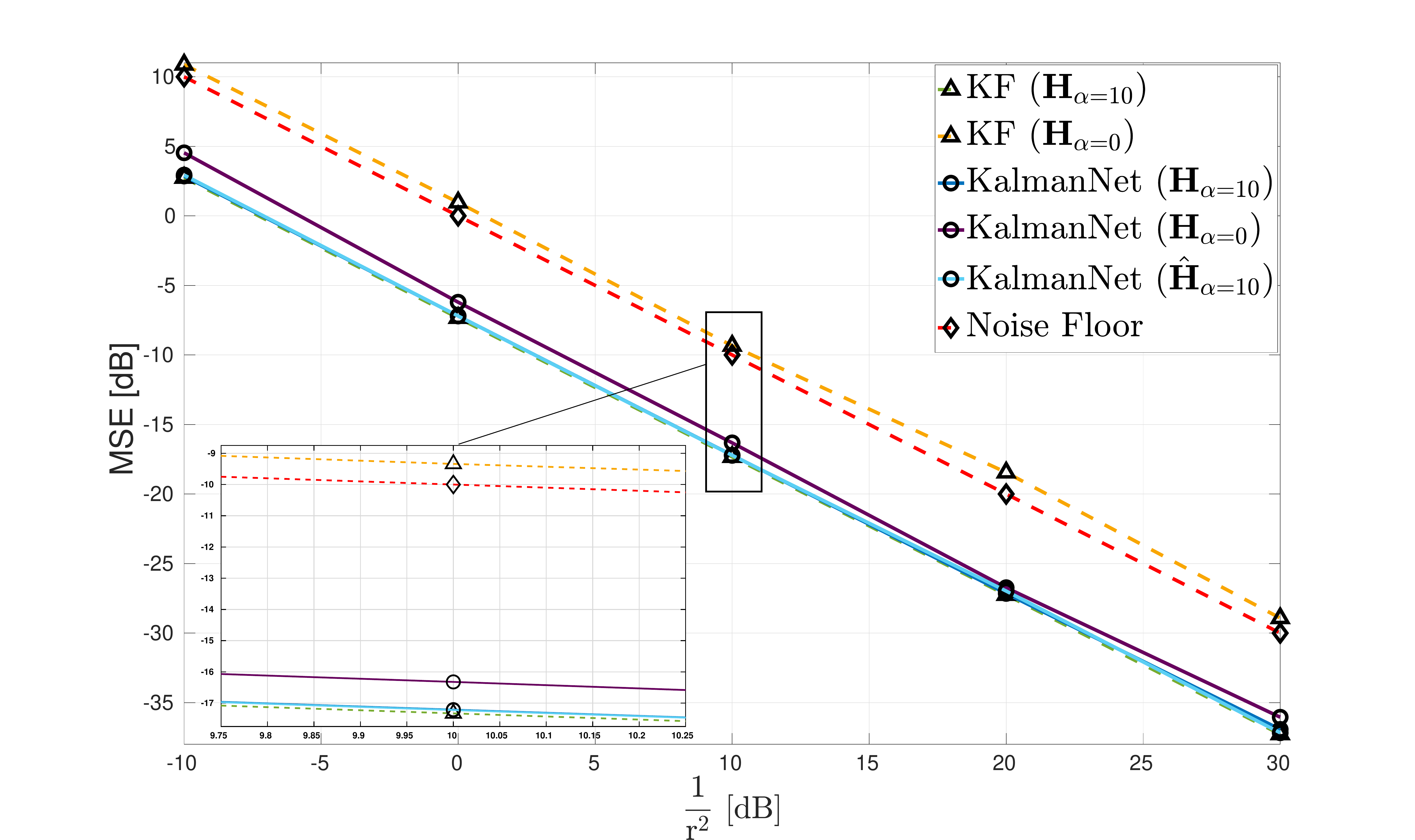}
\caption{State-observation mismatch.}
\label{fig:Lin_obs_mismatch}
\end{subfigure}
\caption{Linear \ac{ss} model, partial information.}
\end{center}
\figSpace
\vspace{-0.2cm}
\end{figure*}

\emph{State-Evolution Mismatch:} Here, we set $T=20$ and $\nu=0\dB$ and use a rotated evolution matrix $\gvec{F}_{\alpha^\circ}, \alpha \in \set{10^\circ, 20^\circ}$ for data generation. The state-evolution matrix available to the filters, denoted $\gvec{F}_0$, is again set to take the controllable canonical form. The mismatched design matrix $\gvec{F}_0$ is related to true $\gvec{F}_{\alpha^\circ}$ via
\begin{equation}\label{eqn:rot_mat_2}
\gvec{F}_{\alpha^\circ}=\gvec{R}^{\gscal{xy}}_{\alpha^\circ}\cdot\gvec{F}_{0}, 
\quad
\gvec{R}^{\gscal{xy}}_{\alpha^\circ}=
\begin{pmatrix}
\cos{\alpha} & -\sin{\alpha}  \\
\sin{\alpha} & \cos{\alpha}
\end{pmatrix}.
\end{equation}
Such scenarios represent a setup in which the analytical approximation of the \ac{ss} model differs from the true generative model. The resulting  \ac{mse} curves  depicted in Fig.~\ref{fig:Lin_evol_mismatch} demonstrate that \acl{kn} (with setting \ref{cfg1a}) achieves a $3\dB$ gain over the \ac{mb} \ac{kf}. In particular, despite the fact that \acl{kn} implements the \ac{kf} with an inaccurate state-evolution model, it learns to apply an alternative \ac{kg}, resulting in \ac{mse} within a minor gap from the \ac{mmse}; i.e., from the \ac{kf} with the true $\gvec{F}_{\alpha^\circ}$ plugged in. 

\emph{\textcolor{NewColor}{State-Observation Mismatch}:} Next, we simulate a setup with state-observation mismatch while setting  $T=100$ and $\nu=-20\dB$. The model mismatch is achieved by using a rotated observation matrix $\gvec{H}_{\alpha=10^\circ}$ for data generation, while using $\gvec{H}=\gvec{I}$ as the observation design matrix. Such scenarios represent a setup in which a slight misalignment $(\approx5\%)$ of the sensors exists. The resulting achieved \ac{mse} depicted in Fig.~\ref{fig:Lin_obs_mismatch} demonstrates that \acl{kn} (with setting \ref{cfg1a}) converges to within a minor gap from the \ac{mmse}. Here, we performed an additional experiment, first estimating the observation matrix from data, and then \acl{kn} used the estimate matrix denoted $\hat{\gvec{H}}_{\alpha}$. In this case it is observed in Fig.~\ref{fig:Lin_obs_mismatch} that \acl{kn} achieves the \ac{mmse} lower bound. 
\textcolor{NewColor}{These results imply that \acl{kn} converges also in distribution to the \ac{kf}}. 
%
%
%
%

%
\subsection{Synthetic Non-Linear Model}
\label{ssec:ToyModel}
\textcolor{NewColor}{Next}, we consider a \acl{nl} \ac{ss} model, where the state-evolution model takes a sinusoidal form, while the state-observation model is a second order polynomial. The resulting \ac{ss} model is given by
\begin{subequations}
\label{eqn:ToyModel}
\begin{align} 
\label{eqn:ToyModela}
\gvec{f}\brackets{\gvec{x}} &= 
\alpha\cdot\sin\brackets{\beta\cdot\gvec{x}+\gvec{\phi}}+\delta,
&
\gvec{x}&\in\greal^2,\\
\label{eqn:ToyModelb}
\gvec{h}\brackets{\gvec{x}} &= 
a\cdot\brackets{b\cdot\gvec{x}+c}^2,
&
\gvec{y}&\in\greal^2.
\end{align}
\end{subequations}
\textcolor{NewColor}{In the following we generate  trajectories of $T=100$ time steps from the noisy \ac{ss} model in \eqref{eq:NL_SS_model}, with $\nu=-20\dB$, while using $\gevol$ and $\gobs$ as in \eqref{eqn:ToyModel} computed in a component-wise manner, with parameters as in  Table~\ref{tbl:NL_TOY_param}. \acl{kn} is used with setting \ref{cfg3}.}

\begin{figure} 
\includegraphics[width=1\columnwidth]{{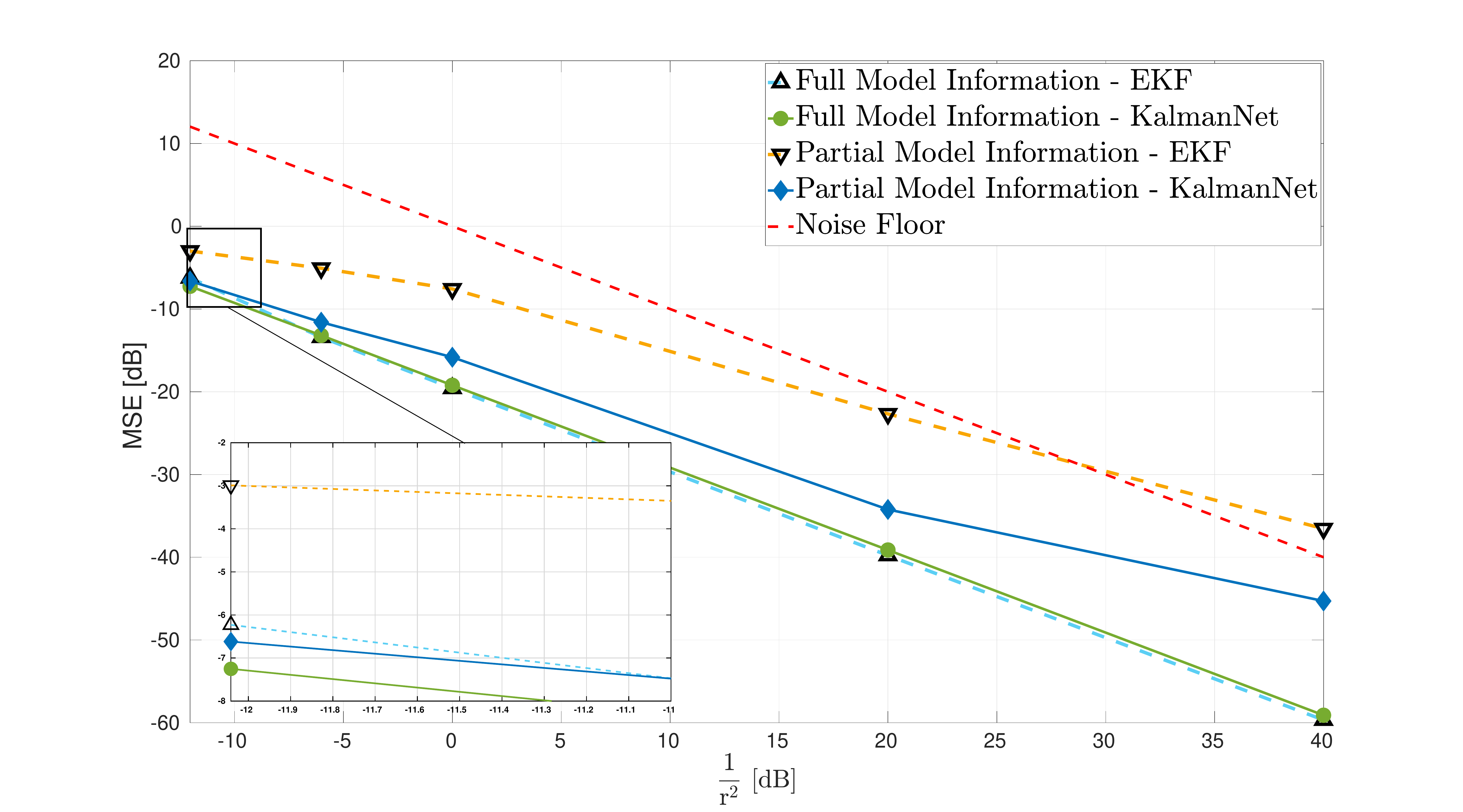}}
\caption{Non-linear \ac{ss} model.
\acl{kn} outperforms \ac{ekf}.}
\label{fig:simToy} 
\figSpace
\end{figure}

\textcolor{NewColor}{
The \ac{mse} values for different levels of observation noise achieved by \acl{kn} compared \textcolor{NewColor}{with} the \ac{mb} \ac{ekf} are depicted in Fig.~\ref{fig:simToy} for both full and partial model information. The full evaluation with the \ac{mb} \ac{ekf}, \ac{ukf}, and \ac{pf} is given in Table~\ref{tbl:NL_sin_full} for the case of full information, and in Table~\ref{tbl:NL_sin_partial} for the case of partial information. We first observe that the \ac{ekf} achieves the lowest \ac{mse} values among the \ac{mb} filters, therefore serving as our main \ac{mb} benchmark in our experimental studies. For full information and in the low noise regime, \ac{ekf} achieves the lowest \ac{mse} values due to its ability to approach the \ac{mmse} in such setups, and \acl{kn} achieves similar performance. For higher noise levels; i.e., for $\gsnr = -12.04$ [dB], the \ac{mb} \ac{ekf} suffers from degraded performance due to a \acl{nl} effect.  Nonetheless, by learning to compute the \ac{kg} from data, \acl{kn} manages to overcome this  and achieves superior \ac{mse}. }

\textcolor{NewColor}{
In the presence of partial model information, the state-evolution parameters used by the filters differs slightly from the true model, resulting in a notable degradation in the performance of the \ac{mb} filters due to the model mismatch. In all experiments, \acl{kn} overcomes such mismatches, and its performance is within a small gap of that achieved when using full information for such setups. We thus conclude that in the presence of harsh non-linearities as well as model uncertainty due to inaccurate approximation of the underlying dynamics, where \ac{mb} variations of the \ac{kf} fail, \acl{kn} learns to approach the \ac{mmse} while maintaining the \acl{rt} operation and low complexity of the \ac{kf}. 
}
%
%
%
%
\begin{table}
\caption{Non-linear toy problem parameters.}
\vspace{-0.3cm}
\begin{center}
{
\begin{tabular}{|c|c|c|c|c|c|c|c| }
\hline
& $\alpha$ & $\beta$ & $\phi$ & $\delta$ & $a$ & $b$ & $c$ \\  
\hline
{Full} & $0.9$ & $1.1$ & $0.1\pi$ & $0.01$ & $1$ & $1$ & $0$ \\  
\hline
{Partial} & $1$ & $1$ & $0$ & $0$ & $1$ & $1$ & $0$ \\
\hline
\end{tabular}
\label{tbl:NL_TOY_param}
}
\end{center}
\figSpace
\end{table}
%
%
\begin{table}
\vspace{0.3cm}
\caption{\ac{mse} $\dB$ – Synthetic \acl{nl} \ac{ss} model; full information.}
\vspace{-0.3cm}
\begin{center}
\scriptsize{
\begin{tabular}{|c|c|c|c|c|c|c|c| }
\hline
\multicolumn{2}{|c|}{ ${1}/{\gscal{r}^2} \dB $} & $-12.04$ & $-6.02$ & $0$ & $20$ & $40$ \\
\hline
\ac{ekf} & $\hat{\mu}$ & -6.23 & \textbf{-13.41} & \textbf{-19.58} & \textbf{-39.78} & \textbf{-59.67}
\\
& $\hat{\sigma}$ & $\pm 0.89$ & $\pm 0.53$ & $\pm 0.47$ & $\pm 0.43$ & $\pm 0.44$
\\
\hline
\ac{ukf} & $\hat{\mu}$ & -6.48 & -13.14 & -18.43 & -27.24 & -37.27
\\
& $\hat{\sigma}$ & $\pm 0.69$ & $\pm 0.49$ & $\pm 0.50$ & $\pm 0.55$ & $\pm 0.31$
\\
\hline
\ac{pf} & $\hat{\mu}$ & -6.59 & -13.33 & -18.78	& -26.70 & -30.98
\\
& $\hat{\sigma}$ & $\pm 0.74$ & $\pm 0.48$ & $\pm 0.39$ & $\pm 0.07$ & $\pm 0.02$
\\
\hline
\acl{kn} & $\hat{\mu}$ &\textbf{-7.25} & -13.19 & -19.22 & -39.13 & -59.10
\\
& $\hat{\sigma}$ & $\pm 0.49$ & $\pm 0.52$ & $\pm 0.55$ & $\pm 0.49$ & $\pm 0.53$
\\
\hline
\end{tabular}
\label{tbl:NL_sin_full}
}
\end{center}
\end{table}
%
%
\begin{table}
\vspace{0.3cm}
\caption{\ac{mse} $\dB$ – Synthetic \acl{nl} \ac{ss} model; partial information.}
\vspace{-0.3cm}
\begin{center}
\scriptsize{
\begin{tabular}{|c|c|c|c|c|c|c|c| }
\hline
\multicolumn{2}{|c|}{ ${1}/{\gscal{r}^2} \dB $} & $-12.04$ & $-6.02$ & $0$ & $20$ & $40$ \\
\hline
\ac{ekf} & $\hat{\mu}$ & -2.99 & -5.07 & -7.57 & -22.67 & -36.55
\\
& $\hat{\sigma}$ & $\pm 0.63$ & $\pm 0.89$ & $\pm 0.45$ & $\pm 0.42$ & $\pm 0.3$
\\
\hline
\ac{ukf} & $\hat{\mu}$ & -0.91 & -1.54 & -5.18 & -24.06 & -37.96
\\
& $\hat{\sigma}$ & $\pm 0.60$ & $\pm 0.23$ & $\pm 0.29$ & $\pm 0.43$ & $\pm 2.21$
\\
\hline
\ac{pf} & $\hat{\mu}$ & -2.32 & -3.29 & -4.83 & -23.66 & -33.13
\\
& $\hat{\sigma}$ & $\pm 0.89$ & $\pm 0.53$ & $\pm 0.64$ & $\pm 0.48$ & $\pm 0.45$
\\
\hline
\acl{kn} & $\hat{\mu}$ &\textbf{-6.62} & \textbf{-11.60} & \textbf{-15.83} & \textbf{-34.23} & \textbf{-45.29}
\\
& $\hat{\sigma}$ & $\pm 0.46$ & $\pm 0.45$ & $\pm 0.44$ & $\pm 0.58$ & $\pm 0.64$
\\
\hline
\end{tabular}
\label{tbl:NL_sin_partial}
}
\end{center}
\end{table}

%
\subsection{Lorenz Attractor} \label{subsec:Lorenz}
The \acl{la} is a three-dimensional chaotic solution to the Lorenz system of ordinary differential equations in \acl{ct}. This synthetically generated system demonstrates the task of online tracking a highly \acl{nl} trajectory and a \acl{rw} practical challenge of handling mismatches due to sampling a {\acl{ct}} signal into \acl{dt}~\cite{gilpin2021chaos}. 

In particular, the noiseless state-evolution of the \acl{ct} process $\gvec{x}_\tau$ with $\tau\in\greal^+$ is given by
\begin{equation}\label{eqn:lorenzDiff}
\frac{\partial}{\partial \tau}{\gvec{x}}_\tau\!=\!
\gvec{A}\brackets{{\gvec{x}}_\tau}\cdot \gvec{x}_\tau,
\hspace{0.075cm}
\gvec{A}\brackets{{\gvec{x}_\tau}}\!=\!
\begin{pmatrix}
-10 & 10 & 0\\
28 & -1 & -x_{1, \tau}\\
0 & x_{1, \tau} & -\frac{8}{3}
\end{pmatrix}.
\end{equation}
To get a \acl{dt}, state-evolution model, we repeat the steps used in \cite{satorras2019combining}. First, we sample the noiseless process with sampling interval $\Delta\tau$ 
%
and assume that $\gvec{A}\brackets{{\gvec{x}}_\tau}$ can be kept constant in a small neighborhood of $\gvec{x}_\tau$; i.e.,
\begin{equation*}
\gvec{A}\brackets{{\gvec{x}}_\tau}\approx    \gvec{A}\brackets{{\gvec{x}}_{\tau+\Delta\tau}}.
\end{equation*}
Then, the \acl{ct} solution of the differential system \eqref{eqn:lorenzDiff}, which is valid in the neighborhood of $\gvec{x}_\tau$ for a short time interval $\Delta\tau$, is
\begin{equation}\label{eqn:ct_Lorenz_sol}
\gvec{x}_{\tau+\Delta\tau}=
\exp\brackets{\gvec{A}\brackets{{\gvec{x}_\tau}}\cdot\Delta\tau}\cdot
\gvec{x}_{\tau}.
\end{equation}
Finally, we take the Taylor series expansion of \eqref{eqn:ct_Lorenz_sol} and a \emph{finite} series approximation (with $J$ coefficients), which results in
\begin{equation}\label{eqn:Taylor}
\gvec{F}\brackets{{\gvec{x}_\tau}}\triangleq
\exp\brackets{\gvec{A}\brackets{{\gvec{x}_\tau}}\cdot\Delta \tau}\approx
\gvec{I} + \sum_{j=1}^{J} \frac{\brackets{\gvec{A}\brackets{{\gvec{x}_\tau}}\cdot\Delta \tau}^j}{j!}.
\end{equation}
The  resulting  \acl{dt} evolution process is given by
\begin{equation}\label{eqn:dt_Lorenz}
\gvec{x}_{t+1} = \gvec{f}\brackets{\gvec{x}_t}=
\gvec{F}\brackets{{\gvec{x}_t}}\cdot\gvec{x}_t.
\end{equation}
The \acl{dt} state-evolution model in \eqref{eqn:dt_Lorenz}, with additional process noise, is used for generating the simulated \acl{la} data. Unless stated otherwise the data was generated with $J=5$ Taylor order and $\Delta \tau = 0.02$ sampling interval. \textcolor{NewColor}{In the following experiments, \acl{kn} is consistently invariant of the distribution of the noise signals, with the models it uses for $\gevol$ and $\gobs$ varying between the different studies, as discussed in the sequel.}
%
%
%
%
\medskip

\subsubsection{Full Information}
We first compare \acl{kn} to the \ac{mb} filter when using the state-evolution matrix $\gvec{F}$ computed via \eqref{eqn:Taylor} with $J=5$. 
\smallskip

\begin{figure*}
\begin{center}
%
%
\begin{subfigure}[pt]{0.99\columnwidth}
\includegraphics[width=1\columnwidth]{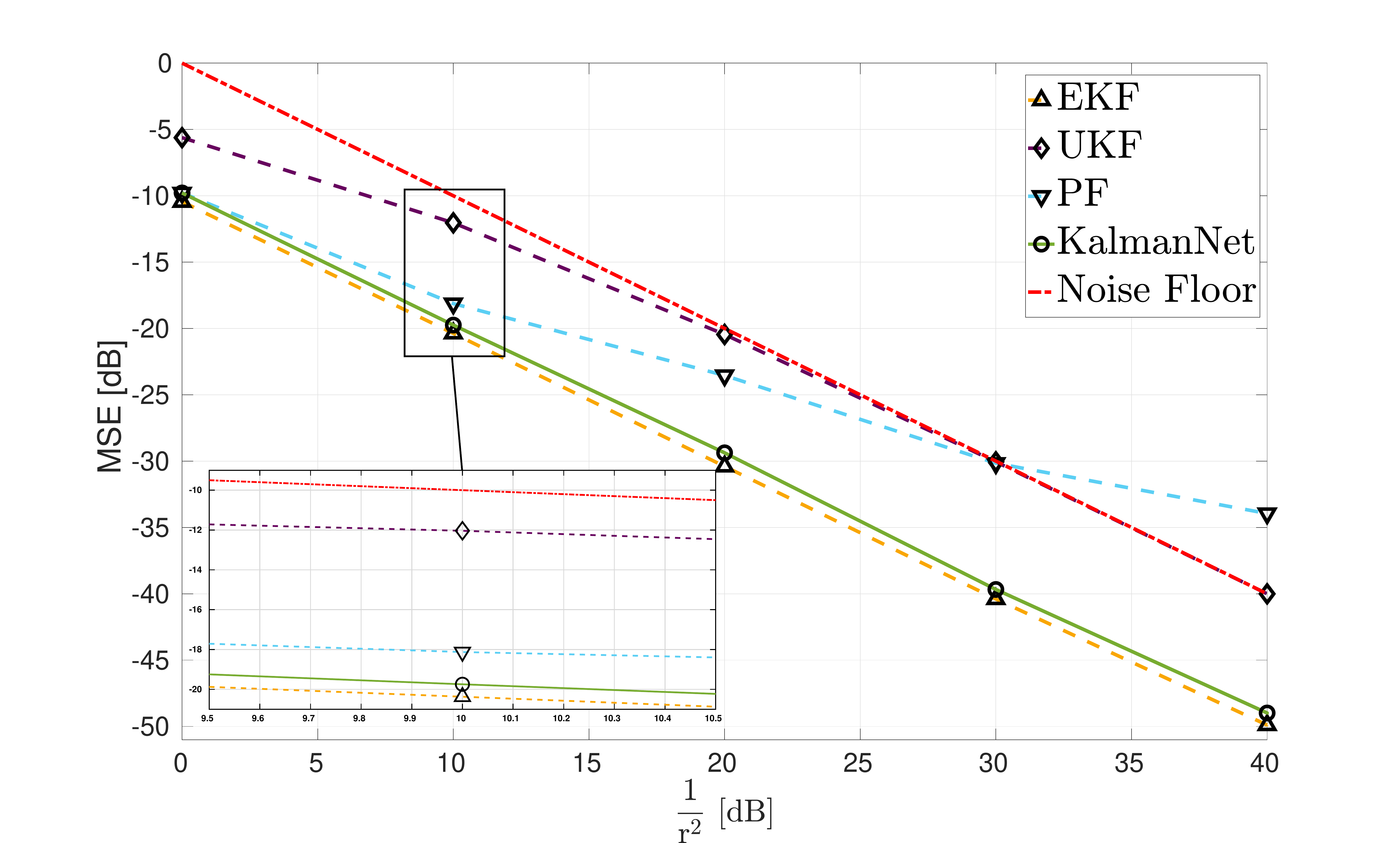}
\caption{$T=2000$, $\nu=-20\dB$, $\gobs=\gvec{I}$.}
\label{fig:sim_Lorenz_full_I}
\end{subfigure}
%
%
\begin{subfigure}[pt]{0.99\columnwidth}
\includegraphics[width=1\columnwidth]{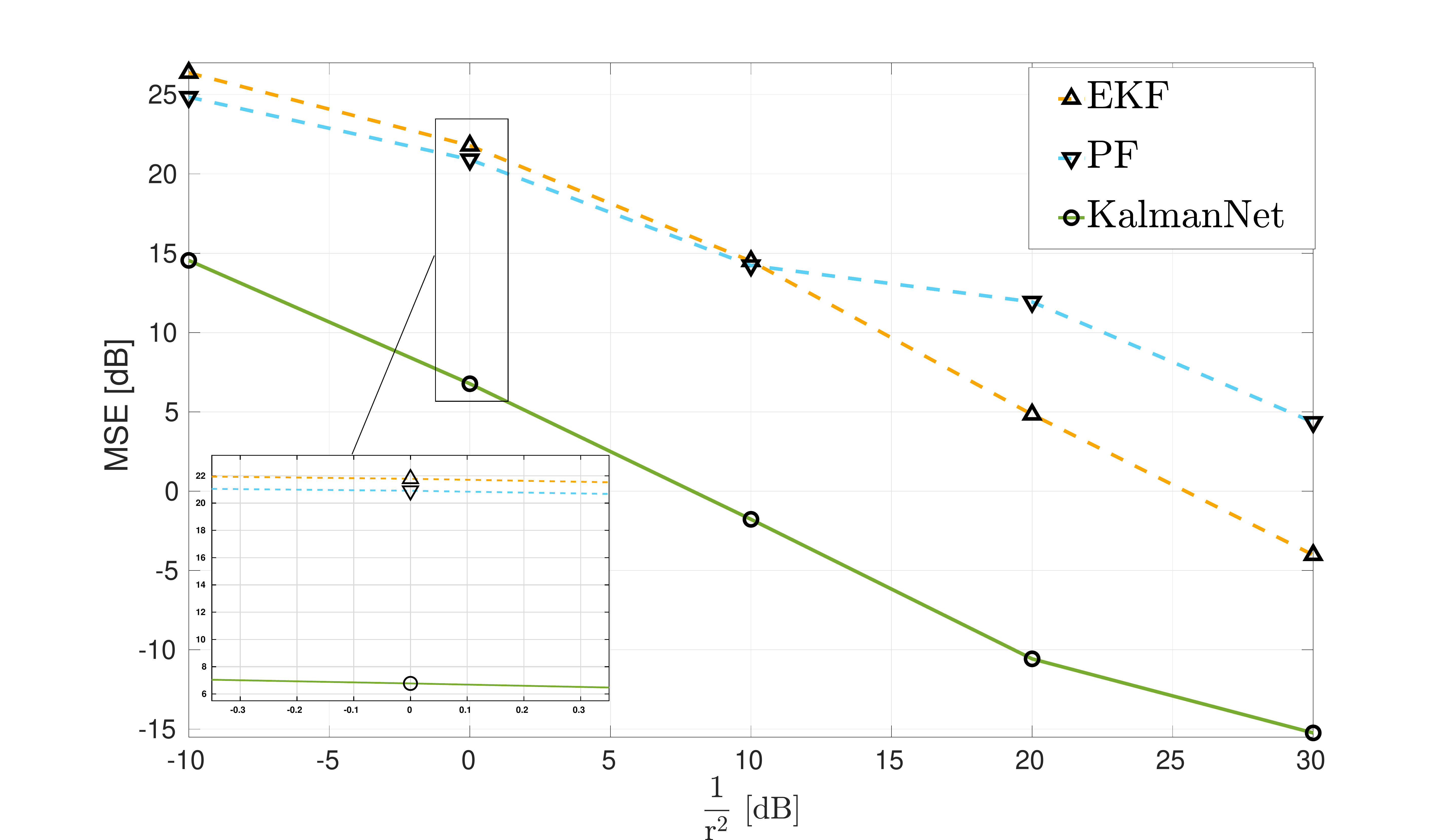}
\caption{$T=20$, $\nu=0\dB$, $\gobs$ \acl{nl}.}
\label{fig:sim_Lorenz_full_NL}
\end{subfigure}
\caption{\acl{la}, full information.}
\label{fig:simLorenz1}
\end{center}
\figSpace
\vspace{-0.2cm}
\end{figure*}

{\bf Noisy state observations}: Here, we set $\gobs$ to be the identity transformation, such that the observations are noisy versions of the true state. Further, we set $\nu=-20\dB$ and $T=2000$. As observed in Fig.~\ref{fig:sim_Lorenz_full_I}, despite being trained on short trajectories $T=100$, \acl{kn} (with setting \ref{cfg2}) achieves excellent \ac{mse} performance—namely, comparable to \ac{ekf}—and outperforms the \ac{ukf} and \ac{pf}. The full details of the experiment are given in Table~\ref{tbl:Lorenz_obs_I}. \textcolor{NewColor}{All the \ac{mb} algorithms were optimized for performance; e.g., applying the \ac{ekf} with full model information achieves an unstable state tracking performance, with \ac{mse} values surpassing $30\dB$. To stabilize the \ac{ekf}, we had to perform a grid search using the available data set to optimize the process noise $\gvec{Q}$ used by the filter.}
%
%
\begin{table}
\vspace{0.3cm}
\caption{\ac{mse} $\dB$ – \acl{la} with noisy state observations.}
\vspace{-0.3cm}
\begin{center}
\footnotesize{
\begin{tabular}{|c|c|c|c|c|c| }
\hline
${1}/{\gscal{r}^2} \dB$ & $0$ & $10$ & $20$ & $30$ & $40$ \\
\hline
\ac{ekf} & \textbf{-10.45} & \textbf{-20.37} & \textbf{-30.40} & \textbf{-40.39} & \textbf{-49.89} \\
\hline
\ac{ukf} & -5.62 & -12.04 & -20.45 & -30.05 & -40.00 \\
\hline
\ac{pf} & -9.78 & -18.13 & -23.54 & -30.16 & -33.95 \\
\hline
\acl{kn} & -9.79 & -19.75 & -29.37 & -39.68 & -48.99 \\
\hline
\end{tabular}
\label{tbl:Lorenz_obs_I}
}
\end{center}
\end{table}
%
%
\smallskip

{\bf {Noisy \acl{nl} observations}}: Next, we consider the case where the observations are given by a non-linear function of the current state, setting $\gvec{h}$ to take the form of a transformation from a cartesian coordinate system to spherical coordinates. We further set $T=20$ and $\nu=0\dB$. From the results depicted in Fig.~\ref{fig:sim_Lorenz_full_NL} and reported in Table~\ref{tbl:Lorenz_obs_NL} we observe that in such non-linear setups, the sub-optimal \ac{mb} approaches operating with full information of the \ac{ss} model are substantially outperformed by \acl{kn} (with setting \ref{cfg3}).
%
%
\begin{table}
\vspace{0.3cm}
\caption{\ac{mse} $\dB$ – \acl{la} with non-linear observations}
\vspace{-0.3cm}
\begin{center}
\footnotesize{
\begin{tabular}{|c|c|c|c|c|c| }
\hline
${1}/{\gscal{r}^2} \dB $& $-10$ & $0$ & $10$ & $20$ & $30$\\
\hline
\ac{ekf} & 26.38 & 21.78 & 14.50 & 4.84 & -4.02\\
\hline
\ac{ukf} & \textrm{nan}	& \textrm{nan} & \textrm{nan} & \textrm{nan} & \textrm{nan} \\
\hline
\ac{pf} & 24.85 & 20.91 & 14.23 & 11.93 & 4.35 \\
\hline
\acl{kn} & \textbf{14.55} & \textbf{6.77} & \textbf{-1.77} & \textbf{-10.57} & \textbf{-15.24} \\
\hline
\end{tabular}
\label{tbl:Lorenz_obs_NL}
}
\end{center}
\end{table} 
%
%
%
%
%
\smallskip

\subsubsection{Partial Information}
W proceed to evaluate \acl{kn} and compare it to its \ac{mb} counterparts under partial model information. We consider \textcolor{NewColor}{three} possible sources of model mismatch arising in the \acl{la} setup: 
\begin{itemize}
\item State-evolution mismatch due to use of a Taylor series approximation of insufficient order.
\item State-observation mismatch as a result of misalignment due to rotation.
\item State-observation mismatch as a result of sampling from \acl{ct} to \acl{dt}.

\end{itemize}
Since the \ac{ekf} produced the best results in the full information case among all \acl{nl} \ac{mb} filtering algorithms, we use it as a baseline for the \ac{mse} lower bound. %
%
%
\smallskip

\begin{figure*}
\begin{center}
%
%
\begin{subfigure}[pt]{0.99\columnwidth}
\includegraphics[width=1\columnwidth]{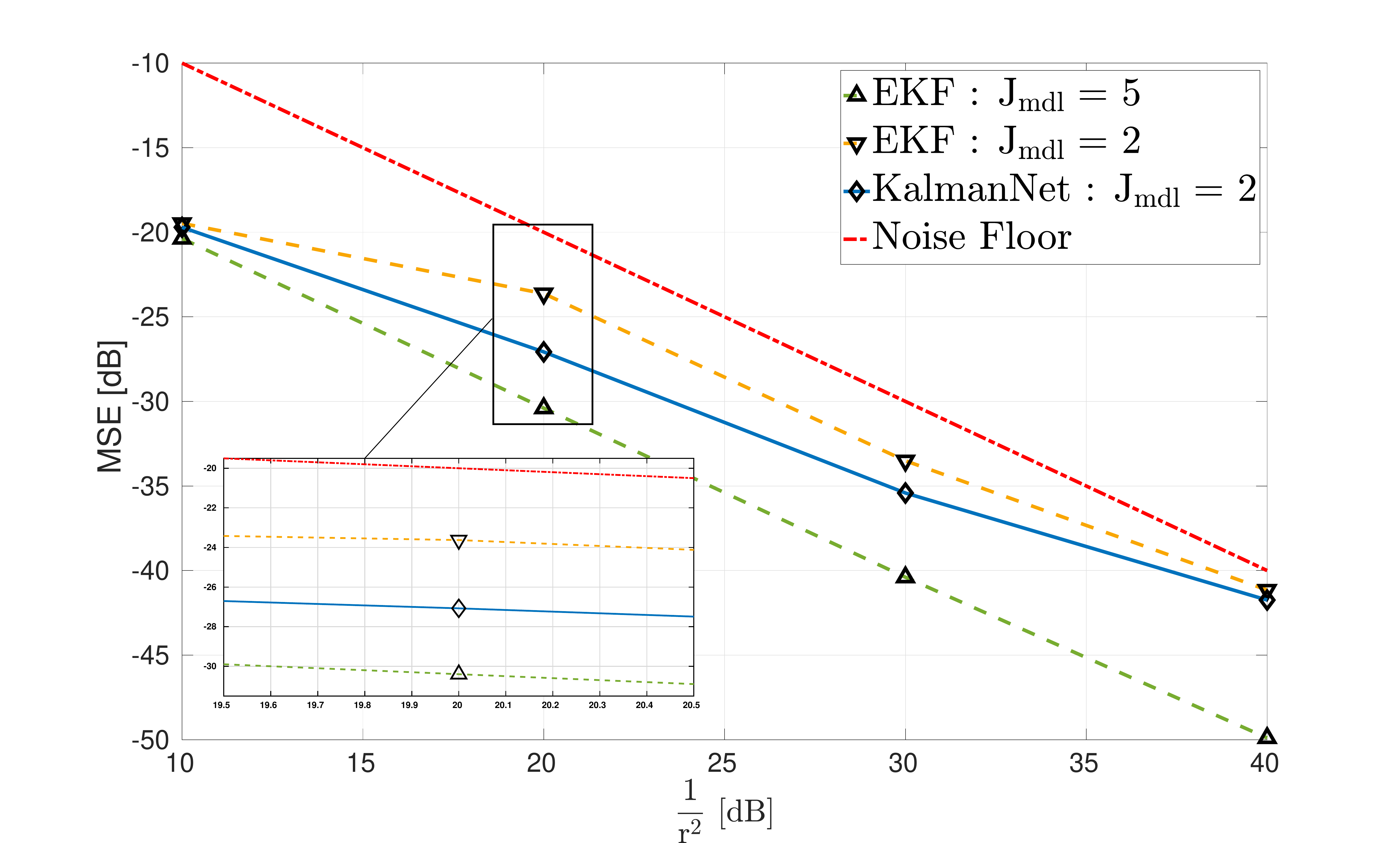}
\caption{State-evolution mismatch, identity $\gvec{h}$, $T=2000$.}
\label{fig:sim_Lorenz_J_2}
\end{subfigure}
%
%
\begin{subfigure}[pt]{0.99\columnwidth}
\includegraphics[width=1\columnwidth]{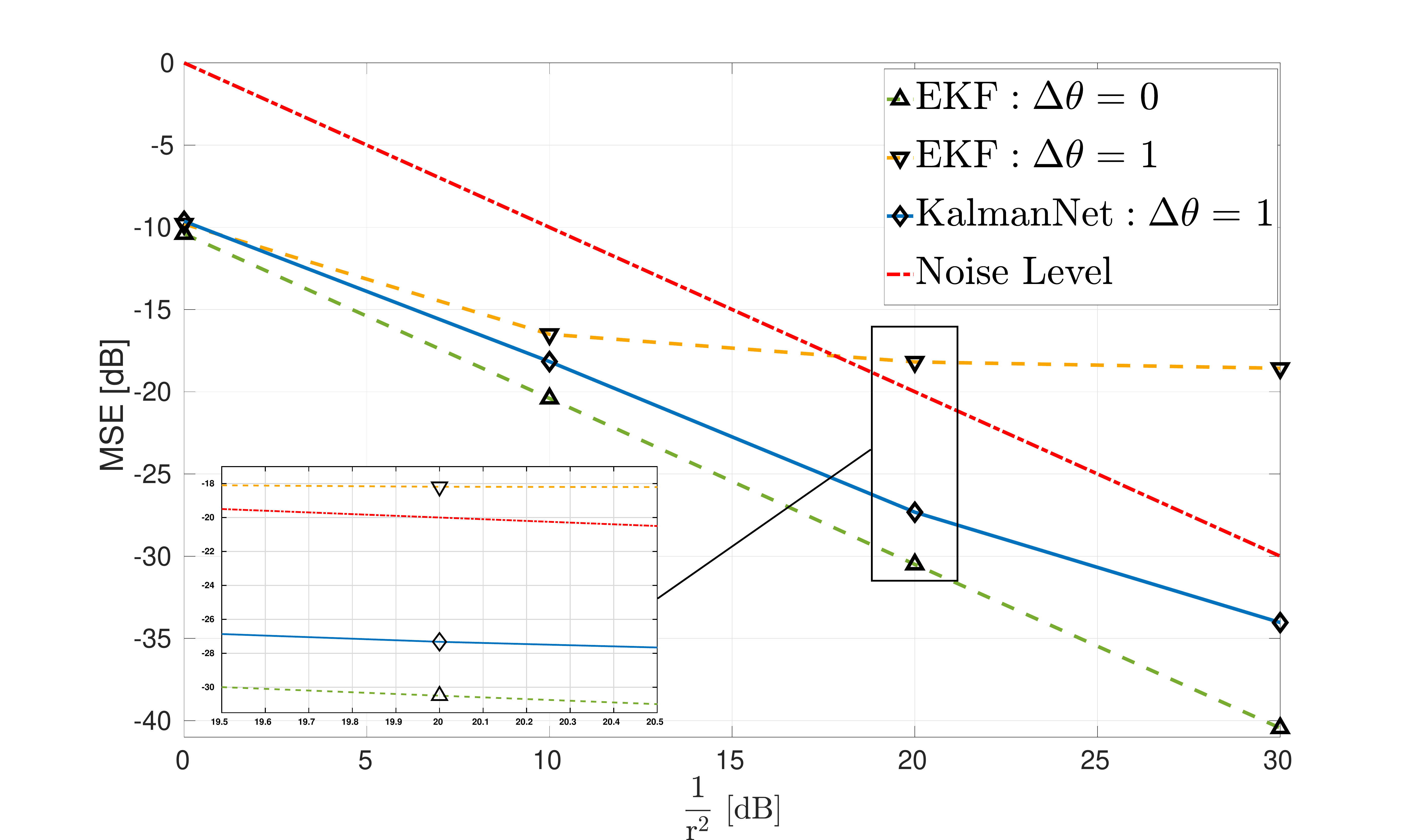}
\caption{Observation mismatch - $\Delta\theta=1^\circ$, \textcolor{NewColor}{$T=1000$}.}
\label{fig:sim_Lorenz_H_rot}
\end{subfigure}
\caption{\acl{la}, partial information.}
\end{center}
\figSpace
\vspace{-0.2cm}
\end{figure*}

{\bf State-evolution mismatch}: 
In this study, both \acl{kn} and the \ac{mb} algorithms operate with a crude approximation of the evolution dynamics obtained by computing \eqref{eqn:Taylor} with $J=2$, while the data is generated with an order $J=5$ Taylor series expansion. We again set $\gvec{h}$ to be the identity mapping, $T=2000$, and $\nu=-20\dB$. The results, depicted in Fig.~\ref{fig:sim_Lorenz_J_2} and reported in Table \ref{tbl:Lorenz_evol_mismatch}, demonstrate that \acl{kn} (with setting \ref{cfg3}) learns to partially overcome this model mismatch, outperforming its \ac{mb} counterparts operating with the same level of partial information. 
%
%
\begin{table}
\vspace{0.3cm}
\caption{\ac{mse} $\dB$ - \acl{la} with  state-evolution mismatch $J=2$.}
\vspace{-0.3cm}
\begin{center}
\footnotesize{
\begin{tabular}{|c|c|c|c|c|c|c| }
\hline
\multicolumn{2}{|c|}{ ${1}/{\gscal{r}^2} \dB $} & $10$ & $20$ & $30$ & $40$\\
\hline
\ac{ekf}   & $\hat{\mu}$ & \textbf{-20.37} & \textbf{-30.40} & \textbf{-40.39} & \textbf{-49.89}\\
 $J=5$&   $\hat{\sigma}$ & $\pm 0.25$ & $\pm 0.24$ & $\pm 0.24$ & $\pm 0.20$\\
\hline
\ac{ekf}  & $\hat{\mu}$ & -19.47 & -23.63 & -33.51 & -41.15\\
  $J=2$&   $\hat{\sigma}$ & $\pm 0.25$ & $\pm 0.11$ & $\pm 0.18$ & $\pm 0.12$\\
\hline
\ac{ukf} & $\hat{\mu}$& -11.95 & -20.45 & -30.05 & -39.98\\
  $J=2$&   $\hat{\sigma}$ & $\pm 0.87$ & $\pm 0.27$ & $\pm 0.09$ & $\pm 0.09$\\
\hline
\ac{pf} & $\hat{\mu}$& -17.95 & -23.47	& -30.11 & -33.81 \\
  $J=2$ &   $\hat{\sigma}$ & $\pm 0.18$ & $\pm 0.09$ & $\pm 0.10$ & $\pm 0.13$\\
\hline
\acl{kn}   & $\hat{\mu}$ & \textbf{-19.71} & \textbf{-27.07} & \textbf{-35.41} & \textbf{-41.74} \\
 $J=2$&   $\hat{\sigma}$ & $\pm 0.29$ & $\pm 0.18$ & $\pm 0.20$ & $\pm 0.11$\\
\hline
\end{tabular}
\label{tbl:Lorenz_evol_mismatch}
}
\end{center}
\end{table}
%
%
\smallskip

{\bf State-observation rotation mismatch}: 
Here, the presence of mismatch in the observations model is simulated by
using data generated by an identity matrix rotated by merely $\theta=1^\circ$. This rotation is equivalent to sensor misalignment of $\approx0.55\%$. The results depicted in Figure.~\ref{fig:sim_Lorenz_H_rot} and reported in Table~\ref{tbl:Lorenz_obs_rot} \textcolor{NewColor}{clearly demonstrate that this seemingly minor rotation can cause a severe performance degradation} for the \ac{mb} filters, while \acl{kn} (with setting \ref{cfg2}) is able to learn from data to overcome such mismatches and to notably outperform its \ac{mb} counterparts, which are sensitive to model uncertainty. Here, we trained \acl{kn} on short trajectories with $T=100$ time steps, tested it on longer trajectories with $T=1000$ time steps, and set $\nu=-20\dB$. This again demonstrates that the learning of \acl{kn} is transferable. 
%
%
\begin{table}
\vspace{0.3cm}
\caption{\ac{mse} $\dB$ - \acl{la} with observation rotation.}
\vspace{-0.3cm}
\begin{center}
\footnotesize{
\begin{tabular}{|c|c|c|c|c|c|}
\hline
\multicolumn{2}{|c|}{ ${1}/{\gscal{r}^2} \dB $}& $0$ & $10$ & $20$ & $30$\\
\hline
\ac{ekf}  & $\hat{\mu}$ &\textbf{-10.40} & \textbf{-20.41} & \textbf{-30.50} & \textbf{-40.45} \\
 $\theta=0^\circ$ &   $\hat{\sigma}$ & $\pm 0.35$ & $\pm 0.37$ & $\pm 0.34$ & $\pm 0.34$\\
\hline
\ac{ekf}  & $\hat{\mu}$ & \textbf{-9.80} & -16.50 & -18.19	& -18.57 \\
 $\theta=1^\circ$ &   $\hat{\sigma}$ & $\pm 0.54$ & $\pm 6.51$ & $\pm 0.22$ & $\pm 0.21$\\
\hline
\ac{ukf} & $\hat{\mu}$ & -2.08 & -6.92 & -7.89 & -8.09 \\
 $\theta=1^\circ$  &   $\hat{\sigma}$ & $\pm 1.73$ & $\pm 0.53$ & $\pm 0.59$ & $\pm 0.62$\\
\hline
\ac{pf}  & $\hat{\mu}$ & -8.48 & -0.18 & 15.24 & 19.87 \\
  $\theta=1^\circ$&   $\hat{\sigma}$ & $\pm 3$ & $\pm 8.21$ & $\pm 3.50$ & $\pm 0.80$\\
\hline
\acl{kn}   & $\hat{\mu}$ & -9.63 & \textbf{-18.17} & \textbf{-27.32} & \textbf{-34.04}\\
 $\theta=1^\circ$&   $\hat{\sigma}$ & $\pm 0.53$ & $\pm 0.42$ & $\pm 0.67$ & $\pm 0.77$\\
\hline
\end{tabular}
\label{tbl:Lorenz_obs_rot}
}
\end{center}
\end{table} 
%
%
%
%
\smallskip

{\bf State-observations sampling mismatch}: 
We conclude our experimental study of the \acl{la} setup with an evaluation of \acl{kn} in the presence of sampling mismatch. Here, we generate data from the \acl{la} \ac{ss} model with an \textcolor{NewColor}{approximate} \acl{ct} evolution process using a dense sampling rate, set to ${\Delta\tau=10^{-5}}$.  We then sub-sample the noiseless observations from the evolution process by a ratio of $\frac{1}{2000}$ and get a decimated process with $\Delta\tau_{\textrm{d}}=0.02$. This procedure results in an inherent mismatch in the \ac{ss} model due to representing an (approximately) \acl{ct} process using a \acl{dt} sequence. In this experiment, no process noise was applied, and the observations are again obtained with $\gvec{h}$ set to identity and \textcolor{NewColor}{$T=3000$}.

\begin{figure*}[t]
\begin{center}
%
%
\begin{subfigure}[pt]{0.99\columnwidth}
\includegraphics[width=1\columnwidth]{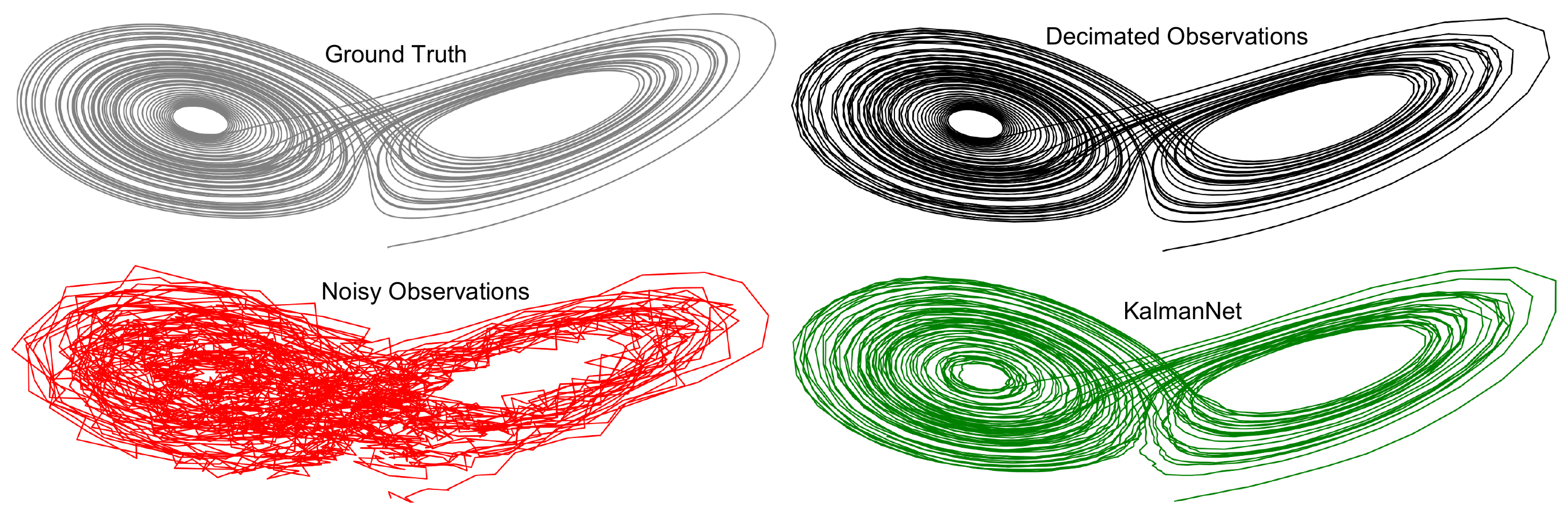}
\end{subfigure}
%
%
\begin{subfigure}[pt]{0.99\columnwidth}
\includegraphics[width=1\columnwidth]{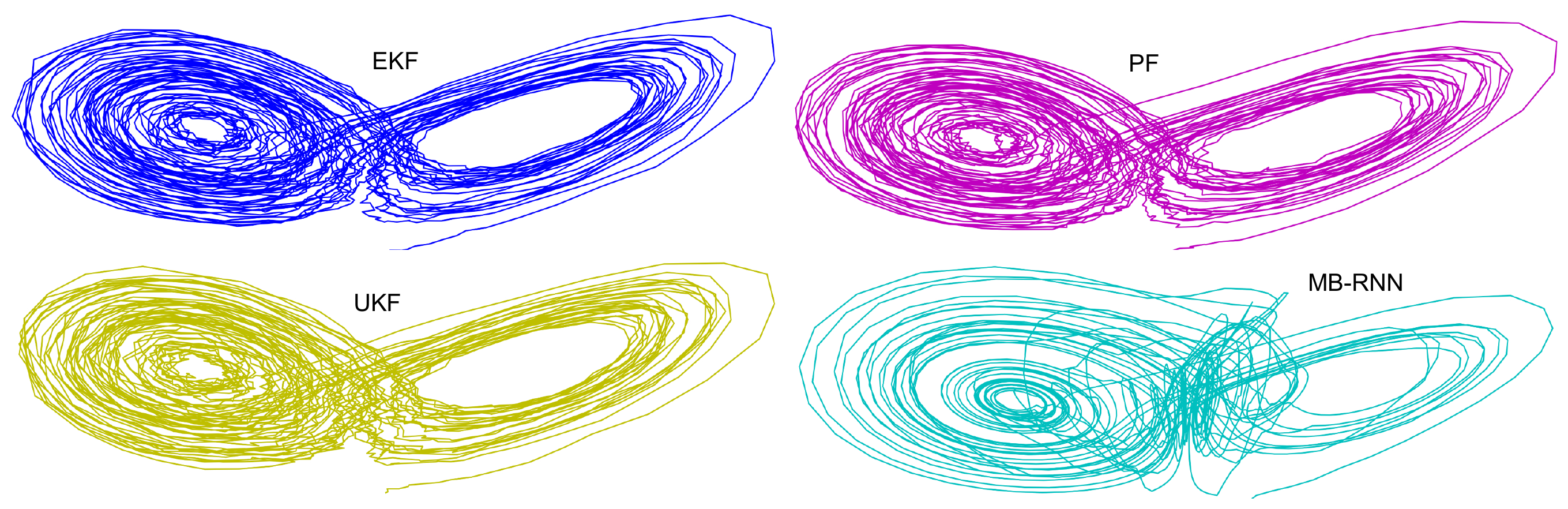}
\end{subfigure}
\caption{\acl{la} with sampling mismatch (decimation), $T=3000$.}
\label{fig:decimation}
\end{center}
\figSpace
\vspace{-0.2cm}
\end{figure*}

The resulting \ac{mse} values for $\frac{1}{\gscal{r}^2}=0\dB$ of \acl{kn} with configuration \ref{cfg3} compared \textcolor{NewColor}{with} the \ac{mb} filters and with the end-to-end neural network termed  \textcolor{NewColor}{\ac{mb}-\ac{rnn}} (see Subsection~\ref{subsec:SynLinear}) are reported in Table~\ref{tbl:decimation}. The results demonstrate that \acl{kn} overcomes the mismatch induced by representing a \acl{ct} \ac{ss} model in \acl{dt}, achieving a substantial processing gain over the \ac{mb} alternatives due to its learning capabilities. \textcolor{NewColor}{The results also demonstrate  that \acl{kn}  significantly outperforms a straightforward combination of domain knowledge; i.e. a state-transition function $\gevol$, with end-to-end \acp{rnn}. A fully model-agnostic \ac{rnn} was shown to diverge when trained for this task}.
In Fig.~\ref{fig:decimation} we visualize how this gain is translated into clearly improved tracking of a single trajectory. To show that these gains of \acl{kn} do not come at the cost of computationally slow inference, we detail the average inference time for all filters (without parallelism). The stopwatch timings were measured on the same platform -- \textit{Google Colab} with CPU: Intel(R) Xeon(R) CPU @ 2.20GHz,  GPU: Tesla P100-PCIE-16GB. We see that \acl{kn} infers faster than the classical methods, thanks to the highly efficient neural network computations and the fact that, unlike the \ac{mb} filters, \textcolor{NewColor}{it does not involve linearization and matrix inversions for each time step.} 

%
%
\begin{table}[t]
\caption{\acl{la} with sampling mismatch.}
\begin{center}
\scriptsize{
\begin{tabular}{|c|c|c|c|c|c|c|}
\hline
$\textrm{Metric}$ &  \ac{ekf}&  \ac{ukf} &   \ac{pf} & \acl{kn} & \ac{mb}-\ac{rnn}\\
\hline
\ac{mse} $\dB$ &  -6.432 & -5.683 & -5.337 & {\bf -11.284} & 17.355\\
 $\hat{\sigma}$ & $\pm 0.093$ & $\pm 0.166$ & $\pm 0.190$ & $\pm 0.301$ & $\pm 0.527$\\
 \hline
 \textrm{Run-time} $[\sec]$ &  5.440
 & 6.072 & 62.946
 & {\bf 4.699} & 2.291\\
 \hline
\end{tabular}
\label{tbl:decimation}
}
\end{center}
\figSpace
\end{table} 
%
%
\subsection{Real World Dynamics: Michigan NCLT Data Set}\label{subsec:NCLT}
In our final experiment we evaluate \acl{kn} on the Michigan \acs{nclt} \acl{ds} \cite{carlevaris2016university}. This \acl{ds} comprises different labeled trajectories, with each one containing noisy sensor readings (e.g., GPS and odometer) and the ground truth \textcolor{NewColor}{locations} of a moving Segway robot. Given these noisy readings, the goal of the tracking algorithm is to localize the Segway from  the raw measurements at any given time. 

To tackle this problem we model the Segway \textcolor{NewColor}{kinematics} (in each axis separately) using the \textcolor{NewColor}{linear} \emph{Wiener} velocity model, where the acceleration is modeled as a white Gaussian noise process \textcolor{NewColor}{$w_\tau$ with variance $\gscal{q}^2$} \cite{bar2004estimation}: 
\begin{equation}\label{eqn:NCLT}
\gvec{x}_\tau=\brackets{p, v}^\top\in\greal^2,
\quad
\frac{\partial}{\partial \tau} {\gvec{x}}_\tau=
\begin{pmatrix}
0 & 1 \\
0 & 0
\end{pmatrix}\cdot\gvec{x}_\tau+
\begin{pmatrix}
0 \\
w_\tau
\end{pmatrix}.    
\end{equation}
Here, $p$ and $v$ are the position and velocity, respectively. The \acl{dt} state-evolution with sampling interval $\Delta \tau$ is approximated as a linear \ac{ss} model in which the evolution matrix $\gvec{F}$ and noise covariance $\gvec{Q}$ are given by
\begin{equation}
\gvec{F} =
\begin{pmatrix}
1 & \Delta \tau\\
0 & 1 
\end{pmatrix},
\quad
\gvec{Q} = \textcolor{NewColor}{\gscal{q}^2}\cdot
\begin{pmatrix}
\frac{1}{3}\cdot\brackets{\Delta \tau}^3 & \frac{1}{2}\cdot\brackets{\Delta \tau}^2 \\
\frac{1}{2}\cdot\brackets{\Delta \tau}^2  & \Delta \tau
\end{pmatrix}.
\end{equation} 
Since \acl{kn} does not rely on  knowledge of the noise covariance matrices, $\gvec{Q}$ is given here for the use of the \ac{mb} \ac{kf} and for completeness. 

The goal is to track the underlying state vector in both axes solely using  odometry data; i.e., the observations are given by noisy velocity readings. In this case the observations obey a noisy linear model: 
\begin{equation}
\gscal{y}\in\greal,
\quad
\gvec{H} =\brackets{0, 1}.
\end{equation}
Such settings where one does not have access to direct measurements for positioning are very challenging yet practical and typical for many applications where \textcolor{NewColor}{positioning} technologies are not available indoors, and one must rely on noisy odometer readings for self-localization. Odometry-based estimated positions typically start drifting away at some point. 

In the assumed model, the x-axis (in cartesian coordinates) are decoupled from the y-axis, and the linear \ac{ss} model used for Kalman filtering is given by
\begin{subequations}
\begin{align}
\tilde{\gvec{F}} &=
\begin{pmatrix}
\gvec{F} & 0\\
0 & \gvec{F}
\end{pmatrix}\in\greal^{4\times4},
\quad
\tilde{\gvec{Q}} =
\begin{pmatrix}
\gvec{Q} & 0\\
0 & \gvec{Q}
\end{pmatrix}\in\greal^{4\times4},\\
\tilde{\gvec{H}} &=
\begin{pmatrix}
\gvec{H} & 0\\
0 & \gvec{H}
\end{pmatrix}\in\greal^{2\times4},
\quad
\tilde{\gvec{R}} =
\begin{pmatrix}
\gscal{r}^2 & 0\\
0 & \gscal{r^2}
\end{pmatrix}\in\greal^{2\times2}.
\end{align}
\end{subequations}
\textcolor{NewColor}{This model is equivalent to applying two independent \acp{kf} in parallel. Unlike the \ac{mb} \ac{kf},  \acl{kn} does not rely on noise modeling, and can thus accommodate  dependency in its learned \ac{kg}.}

We arbitrarily use the session with date 2012-01-22 that consists of a single trajectory. Sampling at $1\textrm{[Hz]}$ results in $5,850$ time steps. We removed unstable readings and were left with 5,556 time steps. The trajectory was split into three sections: $85\%$ for training ($23$ sequences of length $T=200$), $10\%$ for validation ($2$ sequences, $T=200$), and $5\%$ for testing ($1$ sequence, $T=277$). We compare \acl{kn} with setting \ref{cfg1} to end-to-end vanilla \ac{rnn} and the \ac{mb} \ac{kf}, where for the latter the matrices  {$\gvec{Q}$} and $\gvec{R}$ were optimized through a grid search.

\begin{figure}
\centering
\includegraphics[width=1\columnwidth]{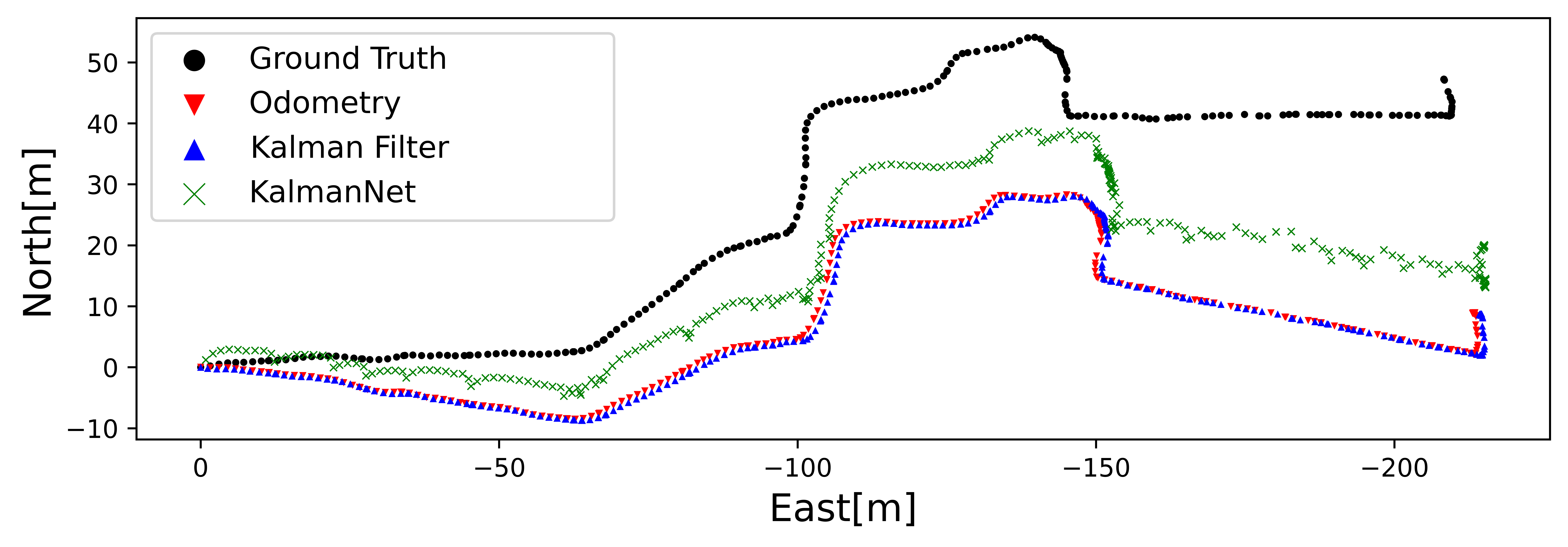}
\caption{NCLT \acl{ds}: ground truth  vs. integrated velocity, trajectory from session with date 2012-01-22 sampled at 1 Hz.} 
\label{fig:NCLT_results}
\end{figure}

Fig. \ref{fig:NCLT_results} and Table~\ref{tbl:NCLT_table} demonstrate the superiority of \acl{kn} for such scenarios. \ac{kf} blindly follows the odometer trajectory and is incapable of accounting for the drift, producing a very similar or even worse estimation than the integrated velocity. The vanilla \ac{rnn}, which is agnostic of the motion model, fails to localize. \acl{kn} overcomes the errors induced by the noisy odometer observations, and provides the most accurate \acl{rt} locations, demonstrating the gains of combining \ac{mb} \ac{kf}-based inference with integrated \ac{dd} modules for \acl{rw} applications. 
%
%
\begin{table}
\caption{Numerical \ac{mse} $\dB$ for the NCLT experiment.}
\begin{center}
{
\begin{tabular}{|c|c|c|c| }
 \hline
 Baseline& \ac{ekf}&   \acl{kn}& Vanilla RNN\\
 \hline
25.47 & 25.385 & {\bf 22.2} & 40.21 \\
 \hline
\end{tabular}
\label{tbl:NCLT_table}
}
\end{center}
\figSpace
\end{table} 
%
%
%
%
%
%
%
\section{Conclusions}\label{sec:Conclusions}
In this work we presented \acl{kn}, a hybrid combination of \acl{dl} with the classic \ac{mb} \ac{ekf}. Our design identifies the \ac{ss}-model-dependent computations of the \ac{mb} \ac{ekf}, replacing them with a dedicated \ac{rnn} operating on specific features encapsulating the information needed for its operation. 
Our numerical study shows that doing so enables \acl{kn} to carry out \acl{rt} \acl{se} in the same manner as  \ac{mb} Kalman filtering, while learning to overcome model mismatches and non-linearities. 
\acl{kn} uses a relatively compact \ac{rnn} that can be trained with a relatively small \acl{ds} and infers a reduced complexity, making it applicable for high dimensional \ac{ss} models and computationally limited devices.
\section*{Acknowledgements}
We would like to thank Prof. Hans-Andrea Loeliger for his helpful comments and discussions, and Jonas E. Mehr for his assistance with the numerical study.
%
%
\bibliographystyle{IEEEtran}
\bibliography{IEEEabrv,KalmanNet}

\begin{thebibliography}{10}
\providecommand{\url}[1]{#1}
\csname url@samestyle\endcsname
\providecommand{\newblock}{\relax}
\providecommand{\bibinfo}[2]{#2}
\providecommand{\BIBentrySTDinterwordspacing}{\spaceskip=0pt\relax}
\providecommand{\BIBentryALTinterwordstretchfactor}{4}
\providecommand{\BIBentryALTinterwordspacing}{\spaceskip=\fontdimen2\font plus
\BIBentryALTinterwordstretchfactor\fontdimen3\font minus
  \fontdimen4\font\relax}
\providecommand{\BIBforeignlanguage}[2]{{%
\expandafter\ifx\csname l@#1\endcsname\relax
\typeout{** WARNING: IEEEtran.bst: No hyphenation pattern has been}%
\typeout{** loaded for the language `#1'. Using the pattern for}%
\typeout{** the default language instead.}%
\else
\language=\csname l@#1\endcsname
\fi
#2}}
\providecommand{\BIBdecl}{\relax}
\BIBdecl

\bibitem{revach2021kalmannet}
G.~Revach, N.~Shlezinger, R.~J.~G. van Sloun, and Y.~C. Eldar, ``{KalmanNet}:
  Data-driven {K}alman filtering,'' in \emph{Proc. IEEE ICASSP}, 2021, pp.
  3905--3909.

\bibitem{durbin2012time}
J.~Durbin and S.~J. Koopman, \emph{Time series analysis by state space
  methods}.\hskip 1em plus 0.5em minus 0.4em\relax Oxford University Press,
  2012.

\bibitem{kalman1960new}
R.~E. Kalman, ``A new approach to linear filtering and prediction problems,''
  \emph{Journal of Basic Engineering}, vol.~82, no.~1, pp. 35--45, 1960.

\bibitem{kalman1961new}
R.~E. Kalman and R.~S. Bucy, ``New results in linear filtering and prediction
  theory,'' 1961.

\bibitem{kalman1963new}
R.~E. Kalman, ``New methods in {W}iener filtering theory,'' 1963.

\bibitem{wiener1949extrapolation}
N.~Wiener, \emph{Extrapolation, interpolation, and smoothing of stationary time
  series: With engineering applications}.\hskip 1em plus 0.5em minus
  0.4em\relax MIT Press Cambridge, MA, 1949, vol.~8.

\bibitem{gruber1967approach}
M.~Gruber, ``An approach to target tracking,'' MIT Lexington Lincoln Lab, Tech.
  Rep., 1967.

\bibitem{larson1967application}
R.~E. Larson, R.~M. Dressler, and R.~S. Ratner, ``Application of the {E}xtended
  {K}alman filter to ballistic trajectory estimation,'' Stanford Research
  Institute, Tech. Rep., 1967.

\bibitem{mclean1962optimal}
J.~D. McLean, S.~F. Schmidt, and L.~A. McGee, \emph{Optimal filtering and
  linear prediction applied to a midcourse navigation system for the
  circumlunar mission}.\hskip 1em plus 0.5em minus 0.4em\relax National
  Aeronautics and Space Administration, 1962.

\bibitem{julier1997new}
S.~J. Julier and J.~K. Uhlmann, ``New extension of the {K}alman filter to
  nonlinear systems,'' in \emph{Signal Processing, Sensor Fusion, and Target
  Recognition VI}, vol. 3068.\hskip 1em plus 0.5em minus 0.4em\relax
  International Society for Optics and Photonics, 1997, pp. 182--193.

\bibitem{gordon1993novel}
N.~J. Gordon, D.~J. Salmond, and A.~F. Smith, ``Novel approach to
  nonlinear/non-{G}aussian {B}ayesian state estimation,'' in \emph{IEE
  proceedings F (radar and signal processing)}, vol. 140, no.~2.\hskip 1em plus
  0.5em minus 0.4em\relax IET, 1993, pp. 107--113.

\bibitem{del1997nonlinear}
P.~Del~Moral, ``Nonlinear filtering: Interacting particle resolution,''
  \emph{Comptes Rendus de l'Acad{\'e}mie des Sciences-Series I-Mathematics},
  vol. 325, no.~6, pp. 653--658, 1997.

\bibitem{liu1998sequential}
J.~S. Liu and R.~Chen, ``Sequential {M}onte {C}arlo methods for dynamic
  systems,'' \emph{Journal of the American Statistical Association}, vol.~93,
  no. 443, pp. 1032--1044, 1998.

\bibitem{auger2013industrial}
F.~Auger, M.~Hilairet, J.~M. Guerrero, E.~Monmasson, T.~Orlowska-Kowalska, and
  S.~Katsura, ``Industrial applications of the {K}alman filter: A review,''
  \emph{{IEEE} Trans. Ind. Electron.}, vol.~60, no.~12, pp. 5458--5471, 2013.

\bibitem{zorzi2016robust}
M.~Zorzi, ``Robust {K}alman filtering under model perturbations,'' \emph{{IEEE}
  Trans. Autom. Control}, vol.~62, no.~6, pp. 2902--2907, 2016.

\bibitem{zorzi2017robustness}
------, ``On the robustness of the {B}ayes and {W}iener estimators under model
  uncertainty,'' \emph{Automatica}, vol.~83, pp. 133--140, 2017.

\bibitem{longhini2021learning}
A.~Longhini, M.~Perbellini, S.~Gottardi, S.~Yi, H.~Liu, and M.~Zorzi,
  ``Learning the tuned liquid damper dynamics by means of a robust {EKF},''
  \emph{arXiv preprint arXiv:2103.03520}, 2021.

\bibitem{lecun2015deep}
Y.~LeCun, Y.~Bengio, and G.~Hinton, ``Deep learning,'' \emph{Nature}, vol. 521,
  no. 7553, p. 436, 2015.

\bibitem{bengio2009learning}
Y.~Bengio, ``Learning deep architectures for {AI},'' \emph{Foundations and
  Trends{\textregistered} in Machine Learning}, vol.~2, no.~1, pp. 1--127,
  2009.

\bibitem{hochreiter1997long}
S.~Hochreiter and J.~Schmidhuber, ``Long short-term memory,'' \emph{Neural
  {C}omputation}, vol.~9, no.~8, pp. 1735--1780, 1997.

\bibitem{chung2014empirical}
J.~Chung, C.~Gulcehre, K.~Cho, and Y.~Bengio, ``Empirical evaluation of gated
  recurrent neural networks on sequence modeling,'' \emph{arXiv preprint
  arXiv:1412.3555}, 2014.

\bibitem{vaswani2017attention}
A.~Vaswani, N.~Shazeer, N.~Parmar, J.~Uszkoreit, L.~Jones, A.~N. Gomez,
  L.~Kaiser, and I.~Polosukhin, ``Attention is all you need,'' \emph{arXiv
  preprint arXiv:1706.03762}, 2017.

\bibitem{zaheer2017latent}
M.~Zaheer, A.~Ahmed, and A.~J. Smola, ``Latent {LSTM} allocation: Joint
  clustering and non-linear dynamic modeling of sequence data,'' in
  \emph{International Conference on Machine Learning}, 2017, pp. 3967--3976.

\bibitem{shlezinger2019viterbinet}
N.~Shlezinger, N.~Farsad, Y.~C. Eldar, and A.~J. Goldsmith, ``{ViterbiNet}: A
  deep learning based {Viterbi} algorithm for symbol detection,'' \emph{{IEEE}
  Trans. Wireless Commun.}, vol.~19, no.~5, pp. 3319--3331, 2020.

\bibitem{shlezinger2019deepSIC}
N.~Shlezinger, R.~Fu, and Y.~C. Eldar, ``{DeepSIC}: Deep soft interference
  cancellation for multiuser {MIMO} detection,'' \emph{{IEEE} Trans. Wireless
  Commun.}, vol.~20, no.~2, pp. 1349--1362, 2021.

\bibitem{shlezinger2020learned}
N.~Shlezinger, N.~Farsad, Y.~C. Eldar, and A.~J. Goldsmith, ``Learned factor
  graphs for inference from stationary time sequences,'' \emph{{IEEE} Trans.
  Signal Process.}, early access, 2022.

\bibitem{shlezinger2020model}
N.~Shlezinger, J.~Whang, Y.~C. Eldar, and A.~G. Dimakis, ``Model-based deep
  learning,'' \emph{arXiv preprint arXiv:2012.08405}, 2020.

\bibitem{carlevaris2016university}
N.~Carlevaris-Bianco, A.~K. Ushani, and R.~M. Eustice, ``University of
  {M}ichigan {N}orth {C}ampus long-term vision and {L}i{DAR} dataset,''
  \emph{The International Journal of Robotics Research}, vol.~35, no.~9, pp.
  1023--1035, 2016.

\bibitem{krishnan2015deep}
R.~G. Krishnan, U.~Shalit, and D.~Sontag, ``Deep {Kalman} filters,''
  \emph{arXiv preprint arXiv:1511.05121}, 2015.

\bibitem{karl2016deep}
M.~Karl, M.~Soelch, J.~Bayer, and P.~Van~der Smagt, ``Deep variational {B}ayes
  filters: Unsupervised learning of state space models from raw data,''
  \emph{arXiv preprint arXiv:1605.06432}, 2016.

\bibitem{fraccaro2017disentangled}
M.~Fraccaro, S.~D. Kamronn, U.~Paquet, and O.~Winther, ``A disentangled
  recognition and nonlinear dynamics model for unsupervised learning,'' in
  \emph{Advances in Neural Information Processing Systems}, 2017.

\bibitem{naesseth2018variational}
C.~Naesseth, S.~Linderman, R.~Ranganath, and D.~Blei, ``Variational sequential
  {M}onte {C}arlo,'' in \emph{International Conference on Artificial
  Intelligence and Statistics}.\hskip 1em plus 0.5em minus 0.4em\relax PMLR,
  2018, pp. 968--977.

\bibitem{archer2015black}
E.~Archer, I.~M. Park, L.~Buesing, J.~Cunningham, and L.~Paninski, ``Black box
  variational inference for state space models,'' \emph{arXiv preprint
  arXiv:1511.07367}, 2015.

\bibitem{krishnan2017structured}
R.~Krishnan, U.~Shalit, and D.~Sontag, ``Structured inference networks for
  nonlinear state space models,'' in \emph{Proceedings of the AAAI Conference
  on Artificial Intelligence}, vol.~31, no.~1, 2017.

\bibitem{satorras2019combining}
V.~G. Satorras, Z.~Akata, and M.~Welling, ``Combining generative and
  discriminative models for hybrid inference,'' in \emph{Advances in Neural
  Information Processing Systems}, 2019, pp. 13\,802--13\,812.

\bibitem{bar2004estimation}
Y.~Bar-Shalom, X.~R. Li, and T.~Kirubarajan, \emph{Estimation with applications
  to tracking and navigation: {T}heory algorithms and software}.\hskip 1em plus
  0.5em minus 0.4em\relax John Wiley \& Sons, 2004.

\bibitem{yuen2016online}
K.-V. Yuen and S.-C. Kuok, ``Online updating and uncertainty quantification
  using nonstationary output-only measurement,'' \emph{Mechanical Systems and
  Signal Processing}, vol.~66, pp. 62--77, 2016.

\bibitem{mu2017stable}
H.-Q. Mu, S.-C. Kuok, and K.-V. Yuen, ``Stable robust {E}xtended {K}alman
  filter,'' \emph{Journal of Aerospace Engineering}, vol.~30, no.~2, p.
  B4016010, 2017.

\bibitem{arasaratnam2007discrete}
I.~Arasaratnam, S.~Haykin, and R.~J. Elliott, ``Discrete-time nonlinear
  filtering algorithms using {G}auss--{H}ermite quadrature,'' \emph{Proc.
  {IEEE}}, vol.~95, no.~5, pp. 953--977, 2007.

\bibitem{arasaratnam2009cubature}
I.~Arasaratnam and S.~Haykin, ``Cubature {K}alman filters,'' \emph{{IEEE}
  Trans. Autom. Control}, vol.~54, no.~6, pp. 1254--1269, 2009.

\bibitem{arulampalam2002tutorial}
M.~S. Arulampalam, S.~Maskell, N.~Gordon, and T.~Clapp, ``A tutorial on
  particle filters for online nonlinear/non-{G}aussian {Bayesian} tracking,''
  \emph{{IEEE} Trans. Signal Process.}, vol.~50, no.~2, pp. 174--188, 2002.

\bibitem{chopin2013smc2}
N.~Chopin, P.~E. Jacob, and O.~Papaspiliopoulos, ``{SMC2}: {A}n efficient
  algorithm for sequential analysis of state space models,'' \emph{Journal of
  the Royal Statistical Society: Series B (Statistical Methodology)}, vol.~75,
  no.~3, pp. 397--426, 2013.

\bibitem{martino2018distributed}
L.~Martino, V.~Elvira, and G.~Camps-Valls, ``Distributed particle
  metropolis-{H}astings schemes,'' in \emph{IEEE Statistical Signal Processing
  Workshop (SSP)}, 2018, pp. 553--557.

\bibitem{andrieu2010particle}
C.~Andrieu, A.~Doucet, and R.~Holenstein, ``Particle {M}arkov chain {M}onte
  {C}arlo methods,'' \emph{Journal of the Royal Statistical Society: Series B
  (Statistical Methodology)}, vol.~72, no.~3, pp. 269--342, 2010.

\bibitem{elfring2021particle}
J.~Elfring, E.~Torta, and R.~van~de Molengraft, ``Particle filters: A hands-on
  tutorial,'' \emph{Sensors}, vol.~21, no.~2, p. 438, 2021.

\bibitem{shumway1982approach}
R.~H. Shumway and D.~S. Stoffer, ``An approach to time series smoothing and
  forecasting using the {EM} algorithm,'' \emph{Journal of {Time Series
  Analysis}}, vol.~3, no.~4, pp. 253--264, 1982.

\bibitem{ghahramani1996parameter}
Z.~Ghahramani and G.~E. Hinton, ``Parameter estimation for linear dynamical
  systems,'' 1996.

\bibitem{dauwels2009expectation}
J.~Dauwels, A.~Eckford, S.~Korl, and H.-A. Loeliger, ``Expectation maximization
  as message passing-part {I}: Principles and {G}aussian messages,''
  \emph{arXiv preprint arXiv:0910.2832}, 2009.

\bibitem{martino2017cooperative}
L.~Martino, J.~Read, V.~Elvira, and F.~Louzada, ``Cooperative parallel particle
  filters for online model selection and applications to urban mobility,''
  \emph{Digital Signal Processing}, vol.~60, pp. 172--185, 2017.

\bibitem{abbeel2005discriminative}
P.~Abbeel, A.~Coates, M.~Montemerlo, A.~Y. Ng, and S.~Thrun, ``Discriminative
  training of {Kalman} filters.'' in \emph{Robotics: Science and Systems},
  vol.~2, 2005, p.~1.

\bibitem{xu2021ekfnet}
L.~Xu and R.~Niu, ``{EKFNet}: Learning system noise statistics from measurement
  data,'' in \emph{Proc. IEEE ICASSP}, 2021, pp. 4560--4564.

\bibitem{barratt2020fitting}
S.~T. Barratt and S.~P. Boyd, ``Fitting a {K}alman smoother to data,'' in
  \emph{IEEE American Control Conference (ACC)}, 2020, pp. 1526--1531.

\bibitem{xie1994robust}
L.~Xie, Y.~C. Soh, and C.~E. De~Souza, ``Robust {K}alman filtering for
  uncertain discrete-time systems,'' \emph{{IEEE} Trans. Autom. Control},
  vol.~39, no.~6, pp. 1310--1314, 1994.

\bibitem{carvalho2010particle}
C.~M. Carvalho, M.~S. Johannes, H.~F. Lopes, and N.~G. Polson, ``Particle
  learning and smoothing,'' \emph{Statistical Science}, vol.~25, no.~1, pp.
  88--106, 2010.

\bibitem{urteaga2016sequential}
I.~Urteaga, M.~F. Bugallo, and P.~M. Djuri{\'c}, ``Sequential {M}onte {C}arlo
  methods under model uncertainty,'' in \emph{IEEE Statistical Signal
  Processing Workshop (SSP)}, 2016, pp. 1--5.

\bibitem{zhou2020kfnet}
L.~Zhou, Z.~Luo, T.~Shen, J.~Zhang, M.~Zhen, Y.~Yao, T.~Fang, and L.~Quan,
  ``{KFNet}: Learning temporal camera relocalization using {K}alman
  filtering,'' in \emph{Proceedings of the IEEE/CVF Conference on Computer
  Vision and Pattern Recognition}, 2020, pp. 4919--4928.

\bibitem{kingma2013auto}
D.~P. Kingma and M.~Welling, ``Auto-encoding variational {B}ayes,'' \emph{arXiv
  preprint arXiv:1312.6114}, 2013.

\bibitem{rezende2014stochastic}
D.~J. Rezende, S.~Mohamed, and D.~Wierstra, ``Stochastic backpropagation and
  approximate inference in deep generative models,'' in \emph{International
  conference on machine learning}.\hskip 1em plus 0.5em minus 0.4em\relax PMLR,
  2014, pp. 1278--1286.

\bibitem{blei2017variational}
D.~M. Blei, A.~Kucukelbir, and J.~D. McAuliffe, ``Variational inference: A
  review for statisticians,'' \emph{Journal of the American Statistical
  Association}, vol. 112, no. 518, pp. 859--877, 2017.

\bibitem{haarnoja2016backprop}
T.~Haarnoja, A.~Ajay, S.~Levine, and P.~Abbeel, ``Backprop kf: Learning
  discriminative deterministic state estimators,'' in \emph{Advances in Neural
  Information Processing Systems}, 2016, pp. 4376--4384.

\bibitem{laufer2018hybrid}
B.~Laufer-Goldshtein, R.~Talmon, and S.~Gannot, ``A hybrid approach for speaker
  tracking based on {TDOA} and data-driven models,'' \emph{{IEEE/ACM} Trans.
  Audio, Speech, Language Process.}, vol.~26, no.~4, pp. 725--735, 2018.

\bibitem{coskun2017long}
H.~Coskun, F.~Achilles, R.~DiPietro, N.~Navab, and F.~Tombari, ``Long
  short-term memory {Kalman} filters: Recurrent neural estimators for pose
  regularization,'' in \emph{Proceedings of the IEEE International Conference
  on Computer Vision}, 2017, pp. 5524--5532.

\bibitem{rangapuram2018deep}
S.~S. Rangapuram, M.~W. Seeger, J.~Gasthaus, L.~Stella, Y.~Wang, and
  T.~Januschowski, ``Deep state space models for time series forecasting,'' in
  \emph{Advances in Neural Information Processing Systems}, 2018, pp.
  7785--7794.

\bibitem{becker2019recurrent}
P.~Becker, H.~Pandya, G.~Gebhardt, C.~Zhao, C.~J. Taylor, and G.~Neumann,
  ``Recurrent {K}alman networks: Factorized inference in high-dimensional deep
  feature spaces,'' in \emph{International Conference on Machine
  Learning}.\hskip 1em plus 0.5em minus 0.4em\relax PMLR, 2019, pp. 544--552.

\bibitem{zheng2017state}
X.~Zheng, M.~Zaheer, A.~Ahmed, Y.~Wang, E.~P. Xing, and A.~J. Smola, ``State
  space {LSTM} models with particle {MCMC} inference,'' \emph{arXiv preprint
  arXiv:1711.11179}, 2017.

\bibitem{salimans2015markov}
T.~Salimans, D.~Kingma, and M.~Welling, ``Markov chain {M}onte {C}arlo and
  variational inference: Bridging the gap,'' in \emph{International Conference
  on Machine Learning}.\hskip 1em plus 0.5em minus 0.4em\relax PMLR, 2015, pp.
  1218--1226.

\bibitem{ni2021rtsnet}
X.~Ni, G.~Revach, N.~Shlezinger, R.~J. van Sloun, and Y.~C. Eldar, ``{RTSNET}:
  Deep learning aided {K}alman smoothing,'' in \emph{Proc. IEEE ICASSP}, 2022.

\bibitem{dey2017gate}
R.~Dey and F.~M. Salem, ``Gate-variants of gated recurrent unit ({GRU}) neural
  networks,'' in \emph{Proc. IEEE MWSCAS}, 2017, pp. 1597--1600.

\bibitem{werbos1990backpropagation}
P.~J. Werbos, ``Backpropagation through time: {W}hat it does and how to do
  it,'' \emph{Proc. {IEEE}}, vol.~78, no.~10, pp. 1550--1560, 1990.

\bibitem{sutskever2013training}
I.~Sutskever, \emph{Training recurrent neural networks}.\hskip 1em plus 0.5em
  minus 0.4em\relax University of Toronto Toronto, Canada, 2013.

\bibitem{klein2021uncertainty}
I.~Klein, G.~Revach, N.~Shlezinger, J.~E. Mehr, R.~J. van Sloun, and Y.~Eldar,
  ``Uncertainty in data-driven {K}alman filtering for partially known
  state-space models,'' in \emph{Proc. IEEE ICASSP}, 2022.

\bibitem{LopezICAS21}
A.~López~Escoriza, G.~Revach, N.~Shlezinger, and R.~J.~G. van Sloun,
  ``Data-driven {K}alman-based velocity estimation for autonomous racing,'' in
  \emph{Proc. IEEE ICAS}, 2021.

\bibitem{rauch1965maximum}
H.~E. Rauch, F.~Tung, and C.~T. Striebel, ``Maximum likelihood estimates of
  linear dynamic systems,'' \emph{AIAA Journal}, vol.~3, no.~8, pp. 1445--1450,
  1965.

\bibitem{kingma2014adam}
D.~P. Kingma and J.~Ba, ``Adam: A method for stochastic optimization,''
  \emph{arXiv preprint arXiv:1412.6980}, 2014.

\bibitem{FilterPy}
\BIBentryALTinterwordspacing
{Labbe, Roger}, \emph{{FilterPy - {Kalman} and {Bayesian} Filters in Python}},
  2020. [Online]. Available: \url{https://filterpy.readthedocs.io/en/latest/}
\BIBentrySTDinterwordspacing

\bibitem{pyParticleEst}
\BIBentryALTinterwordspacing
{Jerker Nordh}, \emph{{pyParticleEst - Particle based methods in Python}},
  2015. [Online]. Available:
  \url{https://pyparticleest.readthedocs.io/en/latest/index.html}
\BIBentrySTDinterwordspacing

\bibitem{gilpin2021chaos}
W.~Gilpin, ``Chaos as an interpretable benchmark for forecasting and
  data-driven modelling,'' \emph{arXiv preprint arXiv:2110.05266}, 2021.

\end{thebibliography}

\end{document}